\documentclass[12pt,a4paper]{article}
\usepackage{type1cm,amsmath,hangcaption,graphicx,indentfirst}
\usepackage[psamsfonts]{amssymb}
\numberwithin{equation}{section}

\begin{document}
\begin{titlepage}

 \renewcommand{\thefootnote}{\fnsymbol{footnote}}
\begin{flushright}
 \begin{tabular}{l}
 KEK-TH-1029\\
 ROM2F/2005/15 \\
 hep-th/0508003\\
 \end{tabular}
\end{flushright}

 \vfill
 \begin{center}
 \font\titlerm=cmr10 scaled\magstep4
 \font\titlei=cmmi10 scaled\magstep4
 \font\titleis=cmmi7 scaled\magstep4
 \centerline{\titlerm D-branes in a Big Bang/Big Crunch Universe:}
 \vskip .3 truecm
 \centerline{\titlerm Misner Space}
 \vskip 2.5 truecm

\noindent{ \large Yasuaki Hikida,$^a$\footnote{E-mail:
hikida@post.kek.jp} Rashmi R.~Nayak$^b$\footnote{E-mail:
Rashmi.Nayak@roma2.infn.it} and Kamal
L.~Panigrahi$^b$\footnote{E-mail: Kamal.Panigrahi@roma2.infn.it}}
\bigskip

 \vskip .6 truecm
\centerline{\it $^a$ Theory Group, High Energy Accelerator Research Organization (KEK)}
\centerline{\it Tukuba, Ibaraki 305-0801, Japan}
\bigskip
\centerline{\it $^b$ Dipartimento di Fisica \& INFN, Sezione di
Roma 2, ``Tor Vergata'',} \centerline{\it Roma  00133, Italy}

 \vskip .4 truecm

 \end{center}

 \vfill
\vskip 0.5 truecm

\begin{abstract}

We study D-branes in a two-dimensional Lorentzian orbifold
${\mathbb R}^{1,1}/ {\Gamma}$ with a discrete boost $\Gamma$.
This space is known as Misner or Milne space, and includes
big crunch/big bang singularity. In this space,
there are D0-branes in spiral orbits and D1-branes with or without
flux on them. In particular, we observe imaginary parts of
partition functions, and interpret them as the rates of 
open string pair creation for D0-branes and 
emission of winding closed strings for D1-branes.
These phenomena occur due to the time-dependence of the background.
Open string $2 \to 2$ scattering amplitude on a D1-brane is also 
computed and found to be less singular than closed string case. 

\end{abstract}
\vfill
\vskip 0.5 truecm

\setcounter{footnote}{0}
\renewcommand{\thefootnote}{\arabic{footnote}}
\end{titlepage}

\newpage

\tableofcontents
\section{Introduction}
\label{Intoduction}

The study of string theory on cosmological backgrounds is of great
interest in view of understanding the stringy phenomena, such as, in 
the early universe cosmology. 
Around the initial space-like
singularity at the beginning of the universe, the field theory
techniques cease to be valid, and the string effects therefore should
be taken into account. However, despite of much progress in string
theory on time-independent backgrounds, this important aspect has
not been widely challenged. This is mainly due to the shortage of
tractable models for strings in cosmological or time-dependent
backgrounds. A few of known ways are to utilize the orbifold
method and the gauged WZW models; A limited list of references on
this subject is
\cite{KOSST,BHKN,CC,Nekrasov,Simon,LMS1,LMS2,CCK,Lawrence,FM,HP}
for the Lorentzian orbifolds and \cite{NW,EGKR,CKR,HT,TT} for the
coset models.

In this paper, we consider a simplest Lorentzian orbifold
${\mathbb R}^{1,1}/\Gamma$, which is two dimensional flat Minkowski
space-time identified by a discrete boost $\Gamma$.
This model describes strings in Misner (Milne) space
\cite{Misner}, which has been studied in
\cite{Nekrasov,BCKR,Pioline1,Pioline2,Pioline3,Pioline4}.
Misner space includes big crunch/big bang singularity
\cite{Misner,HE}, and hence the behavior of strings at the
singularity is of interest. 
However, this space also has regions where closed time-like curves
(CTCs) exist, which might invalidate the theory on this background.
Because of this subtlety, the authors of \cite{Simon,LMS1}
studied the parabolic orbifold ${\mathbb R}^{1,2}/\Gamma$,
which is the three dimensional Minkowski space-time with
the identification of a discrete null boost.
The orbifold admits supersymmetry, light-like Killing vectors
and closed null curves, but no CTC.
Even though these good properties, the orbifold is known to be
unstable due to the large back reaction \cite{Lawrence,LMS2,FM,HP}.
Therefore, it might be enough to study the simpler model to
examine the properties of the big crunch/big bang singularities
in string models. In fact, the Misner space case seems more
interesting, since we can expect particle (string)
productions, which do not occur in the parabolic orbifold
with light-like Killing vectors.

In a Euclidean orbifold, strings in twisted sectors 
play an important
role to resolve the orbifold singularity, thus it is natural to
expect that the big crunch/big bang singularity is also resolved
by condensation of closed strings in twisted sectors.
These twisted strings are investigated in
\cite{Pioline1,Pioline2,Pioline3,Pioline4} by
emphasizing the similarity to open strings in the presence of electric flux, 
though any clear conclusion on the resolution (if any) cannot be made.
This is mainly because the closed strings in twisted sectors
are not localized in the fixed point as seen below.
Interestingly, it was conjectured in \cite{Hagedorn} that
the twisted closed string condensation removes
the regions with CTC as the stringy version of chronology protection
\cite{Hawking}. Recent related works are given in
\cite{YZ,CC2,MS,Horowitz,CSV,BKRS,PPS,KKLN}.

The aim of this paper is to investigate D-branes in Misner space.%
\footnote{This is a work subsequent to \cite{HNP},
where we investigated classical aspects of D-branes in another
big bang/big crunch universe, i.e., D-branes in the Nappi-Witten
gauged WZW model. D-branes in other time-dependent backgrounds are 
investigated, for example, in
\cite{HS,AP,DN,CLO,Okuyama}.}
Since there are D0-branes and D1-branes in the covering space
(two dimensional Minkowski space), there are also D0-branes
and D1-branes in the two dimensional orbifold theory.
In general, D-branes behave very interestingly in curved backgrounds 
contrary to D-branes in flat space. In particular, we observe closed 
string emission from a D-brane falling into the stack of NS5-branes
\cite{Kutasov} using the boundary state formalism
\cite{NST,Sahakyan,CLS,NPRT,NRS}. 
Since our background is time-dependent and includes the
big bang/big crunch singularity, we expect the similar emissions
of strings happen on the D-branes. For the purpose, we compute
the annulus amplitudes for open strings, and examine the emissions 
of strings from the imaginary parts of the amplitudes.
Another interesting observation may be made in
the open string scattering amplitude. It was shown in \cite{BCKR}
that the $2 \to 2$ scattering amplitude diverges due to the graviton
exchange, which is the signal of the large back-reaction.
Since there is no graviton sector in the open string spectrum, it
is worth examining open string scatterings and study 
(would-be) divergences.

The layout of the paper is as follows. In section 2 we review
briefly some aspects of closed strings in Misner space for bosonic
strings. In particular, we examine the spectrum in the twisted
sectors and compute the torus amplitude. We also show the no-ghost
theorem for the closed strings in the twisted sectors. 
In section 3, we first present the effective field theory
analysis on D-branes and write down the classical configurations
of both the D0 and D1-branes. 
We next compute annulus amplitudes for open strings
on a D0-brane and a D1-brane. We also construct the boundary
states for the D-branes and reproduce the amplitudes as
scattering amplitudes between boundary states. 
There are imaginary parts of the annulus amplitudes, 
which indicate the instability of the D-branes. We
see that open strings between image branes are pair-created
on the D0-brane and winding closed strings are emitted from the D1-brane. 
In section 4, we generalize this results to superstring settings,
and check that the emissions are not related to tachyonic states.
In section 5, we turn our attention to the
scattering process, and analyze the divergence structures of
four point functions. We find that the open string four point
function is less singular than closed string one.
In section 6, we investigate open strings and D-branes in 
so-called Grant space \cite{Grant},
which should behave less singular than Misner space case
due to the introduction of an extra non-trivial dimension.  
Finally in section 7 we present the
summary of our result and a set of open questions
deserved to be investigated. We list up the formulae for theta and
eta functions in appendix A.

\section{Closed strings in Misner space}

Misner space can be regarded as a quotient of a two
dimensional flat Minkowski space-time by a Lorentz boost
\begin{align}
x^{\pm} \sim e^{\pm 2\pi \gamma} x^{\pm}
\label{misner}
\end{align}
with $\gamma \in {\mathbb R}$.
The space includes four regions divided by the lines $x^+ x^- = 0$.
Among the four, the two regions $x^+ x^- > 0$ are called as the
cosmological regions.
Using the coordinate transformation
$x^{\pm} = \frac{1}{\sqrt{2}}t e^{\pm \gamma \psi}$,
the metric is given by
\begin{align}
ds^2 = -dt^2 + \gamma^2 t^2 d \psi ^2
\label{cosmologicalr}
\end{align}
with the periodicity $\psi \sim \psi + 2 \pi$.
Therefore, the universe consists of a circle with time-dependent radius.
The universe starts with an infinite large radius and shrinks as time goes.
At $t=0$, the radius becomes zero, where the point may be
interpreted as the big crunch singularity.
Another region starts at $t=0$ with big bang singularity,
expands for a while and ends with an infinite large radius.
There are two other regions $x^+x^- < 0$,
which are called as whisker regions.
The metric in these regions is given by
\begin{align}
ds^2 = d r^2 - \gamma^2 r^2 d \chi^2
\label{whiskerr}
\end{align}
with $x^{\pm} = \pm  \frac{1}{\sqrt{2}} r e^{\pm \gamma \chi}$.
Due to the identification \eqref{misner},
the time coordinate has the periodicity $\chi \sim \chi + 2 \pi$,
and hence the universe includes closed time-like curves (CTCs)
everywhere in the whisker regions.
One may find in \cite{HE} arguments on
more detailed structure of this universe and singularity.

\subsection{Closed strings in the twisted sectors}

Because of the identification \eqref{misner}, the spectrum
includes sectors where
\begin{align}
 X^{\pm} (\tau, \sigma+ 2\pi) = e^{\pm 2\pi
\gamma w} X^{\pm}(\tau, \sigma)
\label{twist}
\end{align}
with $w \in {\mathbb Z}$. Since the spectrum in the untwisted sector
($w=0$) is almost the same as the usual flat case,
we will concentrate on the twisted sectors ($w \neq 0$).
In the Misner space, there are two types of non-trivial cycles;
$\psi$ direction in the cosmological regions and $\chi$ direction
in the whisker regions. Therefore the closed strings may
wrap either of the cycles.

In order to catch the physical picture of winding strings,
we restrict to the lowest modes as
\begin{align}
X^{\pm} (\tau, \sigma) = \pm \sqrt{\frac{\alpha '}{2}} \frac{\alpha^{\pm}_0}{\nu} e^{\pm \nu (\tau + \sigma) }
 \mp \sqrt{\frac{\alpha '}{2}} \frac{\tilde \alpha^{\pm}_0}{\nu} e^{\mp \nu (\tau - \sigma) }
\end{align}
with $\nu= \gamma w$. The Virasoro constraint restricts
the values
\begin{align}
\omega^2 &= \frac{1}{2}(\alpha^+_0 \alpha^-_0 + \alpha^-_0 \alpha^+_0) ~,
& \tilde \omega^2
= \frac12( \tilde \alpha^+_0 \tilde \alpha^-_0 + \tilde \alpha^-_0\tilde
\alpha^+_0)
\end{align}
as
\begin{align}
\omega^2 &= \frac{\alpha '}{4} \vec k ^2 + N - \frac{d}{24} + \frac12 \nu^2 ~,
&\tilde \omega^2 &= \frac{\alpha '}{4} \vec k ^2 + \tilde N - \frac{d}{24} 
 + \frac12 \nu^2 .
\label{tomega}
\end{align}
Here we have included $d$ extra flat dimension, and denoted
$\vec k$ as the momenta and $N,\tilde N$ as the occupation number
operators.
In case $\omega = \tilde \omega$, we may set
\begin{align}
 \alpha^+_0 &= \alpha^-_0 = \epsilon \omega  ~,
 &\tilde \alpha^+_0 &= \tilde \alpha^-_0 = \tilde \epsilon \omega  ~,
\end{align}
where we take $\epsilon,\tilde \epsilon = \pm 1$, though only two choices
are essentially different. If we take $\epsilon = \tilde \epsilon = 1$,
then we obtain
\begin{align}
X^{\pm} (\tau,\sigma) =
 \frac{\sqrt{2 \alpha '}\omega}{\nu} e^{\pm \nu \sigma} \sinh \nu \tau ~.
 \label{short}
\end{align}
This represents a string wrapping $\psi$ direction and localized in
the cosmological regions. Another choice may be
$\epsilon = - \tilde \epsilon = 1$, which leads to
\begin{align}
X^{\pm} (\tau,\sigma ) =
\pm \frac{\sqrt{2 \alpha '}\omega}{\nu} e^{\pm \nu \sigma} \cosh \nu \tau ~.
 \label{long}
\end{align}
This string winds $\chi$ direction and is localized in a whisker
region. 
See \cite{Pioline1,Pioline2,Pioline3,Pioline4}
for physical properties of these winding strings.

\subsection{Torus amplitude}

We re-derive the torus amplitude computed in \cite{CC,Nekrasov}
in order to clarify the way how to compute annulus amplitudes.
First we use the path integral formalism.
We specify the worldsheet coordinate with Euclidean signature as
$(z, \bar z)$ where $z=\sigma_1 + i \sigma_2$ with periodic
boundary conditions $\sigma_1 \sim \sigma_1 + 2 \pi$ and $\sigma_2
\sim \sigma_2 + 2 \pi$. The worldsheet metric is given by
\begin{align}
 ds^2 = (d \sigma_1 + \tau d \sigma_2)
        (d \sigma_1 + \bar \tau d \sigma_2) ~.
\end{align}
Due to the identification \eqref{misner},
we can assign the boundary conditions
\begin{align}
 X^{\pm} (\sigma_1 + 2 \pi ,\sigma_2 )
   &= e^{\pm 2 \pi \gamma w} X^{\pm} (\sigma_1, \sigma_2) ~,
 &X^{\pm} (\sigma_1 , \sigma_2 + 2 \pi )
   &= e^{\pm 2 \pi \gamma k} X^{\pm} (\sigma_1, \sigma_2)
   \label{torusbc}
\end{align}
with $w,k \in {\mathbb Z}$. 
Since the untwisted sector $(w,k)=(0,0)$ is almost the same
as in the flat space case, we focus on the twisted sector
 $(w,k) \neq (0,0)$.
The general solution satisfying the
above boundary conditions is
\begin{align}
 X^{\pm} (\sigma_1,\sigma_2) =
   e^{\pm \gamma (w \sigma_1 + k \sigma_2)}
   \left(\sum_{m,n \in {\mathbb Z}} a_{m,n}^{\pm} e^{i (m \sigma_1 + n \sigma_2 )}\right)
   ~. \label{ef}
\end{align}
Thus the eigenvalue of the Laplacian operator is
\begin{align}
 \Delta  =\frac{1}{\tau_2^2}
 (\partial_2 - \tau \partial_1)(\partial_2 - \bar \tau \partial_1) 
= - \frac{1}{\tau_2^2}
  (n - \tau m + y)(n - \bar \tau m - \bar y)
\end{align}
with $y \,(\equiv y_1 + i y_2)= i \gamma (\tau w - k)$.

Performing integral with respect to $a_{m,n}$, we obtain
\begin{align}
 {\cal T}_{w,k}^{-1} (\tau) = {\rm Det}\, (- \Delta)
  = \prod_{m,n} \frac{1}{\tau_2^2}
  (n - \tau m + y)(n - \bar \tau m - \bar y) ~.
\end{align}
In order to obtain a finite expression,
we use the zeta function regularization.%
\footnote{See, e.g., \cite{Ginsparg,RT}.}
First we find
\begin{align}
 \prod_m \frac{1}{\tau_2^2}
  = \lim_{s \to 0}
   \left( \frac{1}{\tau_2^2 } \right)^{2(\sum_{m \geq 1} m^{-s}) + 1}
  = \left( \frac{1}{\tau_2^2} \right)^{2 \zeta (0) + 1} = 1 ~.
\end{align}
In the last equality, we have used $\zeta (0) = -1/2$.
Next we observe
\begin{align}
 \prod_{m,n} (n - \tau m + y)(n - \bar \tau m - \bar y) 
  = \prod_{n} (n+y)(n - \bar y) \prod_{m\neq 0, n}
      (n - \tau m + y)(n - \bar \tau - \bar y) ~.
\end{align}
Using the formula
\begin{align}
 \prod_{n = -\infty}^{\infty} (n+a)
  = a \prod_{n=1}^{\infty} (-n^2)\left(1-\frac{a^2}{n^2}\right)
  = 2i\sin \pi a ~,
\end{align}
we can compute as
\begin{align}
 \prod_{n} (n+y)(n - \bar y) = 4 \sin \pi y \sin \pi \bar y ~,
\end{align}
and
\begin{align}
&\prod_{m\neq 0 ,n}
      (n - \tau m + y)(n - \bar \tau m - \bar y) \\
      &= \prod_{m \geq 1}
      e^{- \pi i (\tau m - y)} e^{\pi i (\bar \tau m + \bar y)}
      (1-qz^{-1})(1-\bar q \bar z^{-1})
      e^{- \pi i (\tau m + y)} e^{\pi i (\bar \tau m - \bar y)}
      (1-qz)(1-\bar q \bar z) ~. \nonumber
\end{align}
Here we have denoted as $q=\exp (2 \pi i \tau)$ with 
$\tau = \tau_1 + i \tau_2$ and $z=\exp(2 \pi i y)$.
Applying the generalized zeta function regularization
\begin{align}
 \sum_{n=1}^{\infty}  (n+\theta)
 = \lim_{s \to -1} \sum_{n=1}^{\infty} (n + \theta)^{-s}
 = - \frac{1}{12} + \frac{1}{2} \theta (1 - \theta) ~,
\end{align}
we obtain
\begin{align}
 \prod_{m \geq 1}
 e^{- \pi i (\tau m - y)} e^{\pi i (\bar \tau m + \bar y)}
  = e^{2 \pi \tau_2 \sum_{m \geq 1} (m + \frac{ i y_1}{\tau_2})}
  = e^{-\frac{\pi \tau_2}{6}+ i \pi y_1
       + \frac{\pi y_1^2}{\tau_2}} ~.
\end{align}
Combining everything, we reproduce the partition function given
in \cite{Nekrasov} as
\begin{align} 
 {\cal T}_{w,k}(\tau) = {\rm Det}^{-1} (- \Delta) =
  \left|\frac{\eta(\tau)}{\vartheta_1( y |\tau)}\right|^2
   e^{-\frac{2\pi y_1^2}{\tau_2}} ~.
   \label{toruspi}
\end{align}
Summing over all twisted sectors $(w,k) \neq (0,0)$,%
\footnote{The full torus amplitude includes the untwisted sector,
and the contribution from the sector is modular invariant by itself.}
we obtain the modular invariant partition function under
$\tau \to \tau + 1$ and $\tau \to -1/\tau$.
Notice that the winding numbers are interchanged $(w,k) \to (k,-w)$
under the S-transformation $\tau \to - 1/\tau$.

This result can be obtained also in the oscillator formalism.
In the $w$-th twisted sector \eqref{twist}, the mode
expansion is given by
\begin{align}
 X^{\pm} (\tau,\sigma) = i \sqrt{\frac{\alpha '}{2}} \sum_n \left[
  \frac{\alpha^{\pm}_n}{n \pm i\nu} e^{-i(n \pm i\nu)(\tau + \sigma)} +
  \frac{\tilde \alpha^{\pm}_n}{n \mp i\nu} e^{-i(n \mp i\nu)(\tau - \sigma)}
  \right]
\end{align}
with $n \in {\mathbb Z}$ and $\nu = \gamma w$. 
The oscillators satisfy the following
commutation relations
\begin{align}
 [\alpha ^+_m,\alpha ^-_n] &= (- m - i \nu )\delta_{m+n} ~,
&[\tilde \alpha ^+_m, \tilde \alpha ^-_n] &= (- m + i \nu )\delta_{m+n} ~.
\end{align}
In terms of oscillators, the Virasoro generators are written as
\begin{align}
\begin{aligned}
 L_0 &= \frac12 i \nu (1 - i \nu)
      - \sum_{n \geq 1} \alpha^+_{-n}  \alpha^-_n
      - \sum_{n \geq 0} \alpha^-_{-n}  \alpha^+_n ~, \\
 \tilde L_0 &= \frac12 i \nu (1 - i \nu)
      - \sum_{n \geq 0} \tilde \alpha^+_{-n}  \tilde \alpha^-_n
      - \sum_{n \geq 1} \tilde \alpha^-_{-n}  \tilde \alpha^+_n
\end{aligned}
      \label{virasorot}
\end{align}
for the Misner space part. As in \cite{CC,Nekrasov},
we have set $\alpha^-_0$ and $\tilde \alpha^+_0$ as
creation operators and $\alpha^+_0$ and $\tilde \alpha^-_0$
as annihilation operators.%
\footnote{For $\nu < 0$, we set $\alpha^+_0$ and $\tilde \alpha^-_0$ as
creation operators and $\alpha^-_0$ and $\tilde \alpha^+_0$
as annihilation operators oppositely to $\nu > 0$ case.
The following equations are modified accordingly.
}\footnote{%
This definition of Hilbert space is actually different from
the previous one, and it is interpreted as a Wick rotation. 
In fact, we can reproduce the torus
amplitude computed in the path integral as shown below.
The relation will be discussed in more detail in the next subsection.
See also \cite{Pioline1}.}

When computing the partition function,
it is convenient to define the occupation
number operators $ N, \tilde N$
 and the boost operators $ J, \tilde J$ as
\begin{align}
\begin{aligned}
 N &= \sum_{n \geq 1} (n N_n^+ + n N_n^-) ~,
 &\tilde N &= \sum_{n \geq 1} (n \tilde N_n^+ + n \tilde N_n^-) ~,
 \\
 i J&= \sum_{n\geq 1} N^+_n - \sum_{n \geq 0} N^-_n ~,
 &i \tilde  J&= \sum_{n\geq 0} \tilde N^+_n
            - \sum_{n \geq 1} \tilde N^-_n ~,
\end{aligned}
\end{align}
where the oscillator number operators are
\begin{align}
 N^{\pm}_n &= - \frac{1}{n\mp i\nu} \alpha^{\pm}_{-n}\alpha^{\mp}_n~,
 &\tilde N^{\pm}_n &= - \frac{1}{n\pm i\nu}
       \tilde \alpha^{\pm}_{-n} \tilde \alpha^{\mp}_n ~.
\end{align}
Then we can reproduce the partition function \eqref{toruspi} by
\begin{align}
 {\cal T}(\tau ) &= \sum_{ \{ k,w \} \neq \{ 0,0 \}} {\cal T}_{k,w} (\tau )~, \nonumber \\
 {\cal T}_{k,w} (\tau )&= {\rm Tr}_w g^k q^{L_0 - \frac{1}{12}}
  \bar q^{\tilde L_0 - \frac{1}{12}}
   = (q \bar q)^{-\frac{1}{12} + \frac{1}{2} i \nu ( 1-i\nu )}
       {\rm Tr}_w e^{ 2 \pi \gamma k i(J + \tilde J)} q^{\nu J + N}
          \bar q^{-\nu \tilde J + \tilde N} ~.
\end{align}
The trace Tr$_w$ is taken over the Hilbert space of $w$-th twisted sector.
The projection operator $\sum_k g^k$ with  
$g = \exp (2 \pi i \gamma (J + \tilde J))$ is needed to
project the Hilbert space into subspace invariant under
the discrete boost \eqref{misner}.
Notice that the twist number $k$ corresponds to the winding number
$k$ along $\sigma_2$ cycle in the path integral formalism
(see \eqref{ef}).

In order to obtain the expression in a critical string
theory, we add
contributions from extra $d$ flat directions
($d=24$ for bosonic string theory) and ghost part.
Then, we obtain the partition function as%
\footnote{We will neglect volume factors, such as, $iV_d$ in the
case of $d$ dimensional Minkowski space-time.}
\begin{align}
{\cal T}
 = \sum_{k,w} \int_{\cal F} \frac{d^2 \tau}{4 \pi^2 \alpha ' \tau_2^2}
  \frac{ e^{-2\pi \tau_2 \gamma^2 w^2} }
   {(4\pi^2 \alpha ' \tau_2)^{\frac{d-2}{2}}|\vartheta_1 (i\gamma (w \tau +k)|\tau) \eta(\tau )^{d-3}|^{2}}  ~,
\label{1-loop}
\end{align}
where ${\cal F}$ denotes the fundamental region
($ | \tau_1 | < \frac12 , |\tau| > 1$).

\subsection{No-ghost theorem}
\label{noghost}

In this subsection, we show no-ghost theorem
for the spectrum of closed strings in the $w$-th twisted sector.
For simplicity, we consider critical bosonic string theory
with $d=24$, and focus on the left-moving part.%
\footnote{The left-moving part of the spectrum of 
twisted closed string is almost the same as that of 
open strings between D-branes with different electric flux.}
We define the Hilbert space by the action of the creation modes
$\alpha^{\pm}_{-n}$, $\alpha^i_{-n}$ with $n \geq 1$ and
ghost modes $c_{-n}$ with $n \geq 0$ and $b_{-n}$ with
$n \geq 1$. 
We use the vacuum state satisfying
\begin{align}
 \frac{1}{2} (\alpha^+_0 \alpha^-_0 + \alpha^-_0 \alpha^+_0)
  | \omega^2 , \vec k \rangle &= \omega^2  | \omega^2 , \vec k \rangle ~,
&\vec \alpha_0 | \omega^2 , \vec k \rangle
   &= \sqrt{\frac{\alpha '}{2}} \vec k | \omega^2 , \vec k \rangle ~,
& b_0 | \omega^2 , \vec k \rangle &= 0
\label{vacuum}
\end{align}
with $\omega^2 \in {\mathbb C}$.
An important property of the vacuum state is
\begin{align}
 \alpha^{\pm}_0 | \omega^2 , \vec k \rangle
 \propto | \omega^2 \pm i \nu , \vec k \rangle ~,
 \label{shiftomega}
\end{align}
which means that the action of the zero modes is just the shift of
the eigenvalue $\omega^2$.
The physical states are then constructed as the cohomology of the
BRST operator
\begin{align}
 Q_B = \sum_n c_n L_{-n} + \sum_{m,n} \frac{(m-n)}{2}
  : c_m c_n b_{-m-n} : - c_0 ~.
\end{align}
In other words, the physical states satisfy $Q_B \psi = 0$ with
equivalent relation $\psi \sim \psi + Q_B \chi$.
In order to show the no-ghost theorem, we will see
that the physical Hilbert space is isomorphic to
${\cal H}^{\perp}$ which has no ghost and light-cone
excitations.

It is convenient to introduce filtration degree counted by%
\footnote{We mainly follow the analysis in \cite{KO,Polchinski}.
For discussions on no-ghost theorem for related
models, see, e.g., \cite{AN1,BHKN,AN2}.}
\begin{align}
 N_f = - \sum_{n \neq 0} \frac{1}{n + i \nu} \alpha^-_{-n} \alpha^+_{n}
       + \sum_{n=1}^{\infty} (c_{-n} b_n - b_{-n} c_n) ~,
\end{align}
then we can split the BRST operator by the degree
\begin{align}
Q_B &= Q_0 + Q_1 + Q_2 ~,
& [ N_f , Q_j ] &= j Q_j ~.
\end{align}
As shown in  \cite{BT,BMP}, non-trivial cohomology of $Q_0$ is
isomorphic to non-trivial cohomology of $Q_B$.
Therefore, it is enough for our purpose to show that the cohomology
of $Q_0 \,(= \sum_{n \neq 0} c_n \alpha^+_{-n} \alpha_0^-)$
is isomorphic to ${\cal H}^{\perp}$.
As in \cite{KO,Polchinski}, we define
\begin{align}
 R = \sum_{n \neq 0} b_n \alpha^-_{-n} ~,
\end{align}
which gives
\begin{align}
 S = \{ Q_0 , R \} = - 
 \sum_{n \geq 1} \left\{ (n - i \nu )
 [ N_n^- + b_{-n} c_n ] + ( n + i \nu ) [ N_n^+ + c_{-n} b_n ]
 \right\} \alpha^-_0 ~.
\end{align}
If $S | \psi \rangle = s \alpha^-_0 | \psi \rangle$ with $s \neq 0$,
then we have
\begin{align}
|\psi ' \rangle \, ( \equiv \alpha^-_0 | \psi \rangle)
 =  \frac{1}{s} \{ Q_0 ,R \} | \psi \rangle
 =  \frac{1}{s} Q_0 R | \psi \rangle
\end{align}
for $Q_0 | \psi \rangle = 0$ and hence $Q_0 | \psi ' \rangle = 0$.
Therefore, non-trivial cohomology exists only if $s=0$,
which means that the non-trivial cohomology is isomorphic to
${\cal H}^{\perp}$.
A crucial difference from the usual flat space case is that
now the zero modes $\alpha^{\pm}_0$ do not commute with each other.
However, thanks to the definition of vacuum \eqref{vacuum}
and its property \eqref{shiftomega}, we can mimic the analysis
in flat space case \cite{KO,Polchinski} as seen above.

Next, let us see this observation is consistent with the one-loop
amplitude \eqref{1-loop}.
In order to do so, we will use the identity
\begin{align}
 \frac{1}{2 \sin T}
 = i \sum_{n=0}^{\infty} e^{-i(2n+1) T}
 = \int_{- \infty}^{\infty} d (\omega^2) \rho (\omega^2 ) e^{- 2 \omega^2 T}
\label{identity}
\end{align}
where
\begin{align}
 \rho (\omega^2 ) = \frac12 (1 + \tanh \pi \omega^2 ) ~.
\end{align}
This equality is valid only when $0 < T < 1$, but we assume
this is true also for the other $T$.\footnote{Notice that we use a
different $\rho (\omega^2)$ from \cite{Pioline1} such that there is a
valid region for real $T$. The difference may originate
from different regularization schemes.}
The most important property of $\rho (\omega^2)$ is
that it is invariant under
$\omega ^2 \to \omega ^2 + i n$ with $n \in {\mathbb Z}$.
Using the identity \eqref{identity} with 
$T=\pi ( i \gamma ( w \tau + k))$ and 
$\rho (\omega^2 + i n) = \rho (\omega ^2)$,
the one-loop amplitude can be rewritten as
\begin{align}
&{\cal T} = \sum_{k,w} \int_{\cal F} \frac{d^2 \tau}{\tau_2 }
 \int \frac{d^{24} k}{(2 \pi)^{24}}
 e^{- \pi \tau_2 \alpha ' \vec k^2 - 2 \pi \tau_2 \gamma ^2 w^2}
  \left| \int d (\omega^2) \rho (\omega^2)
 \sum_{states} e^{2 \pi i \gamma k ( J - \omega ^2 )}
 q^{ \nu (J - \omega ^2) + N - 1 } \right|^2 \nonumber \\
 & \quad =
 \sum_{k,w} \int_{\cal F} \frac{d^2 \tau}{\tau_2}
 \int \frac{d^{24} k}{(2 \pi)^{24}}
 e^{- \pi \tau_2 \alpha ' \vec k^2 - 2 \pi \tau_2 \gamma ^2 w^2}
  \left| \frac{1}{2 \sin \pi ( i \gamma ( w \tau + k))}
  \sum_{states \in {\cal H}^{\perp}} q^{N - 1 } \right|^2 ~.
\end{align}
The last expression implies that negative norm states do not propagate
in the one-loop amplitude.

Here we should comment on the interpretation of \eqref{identity}.
As seen in the torus amplitude,
the left hand side of \eqref{identity} may be interpreted
as the contribution of zero modes, which are
treated as creation and annihilation modes. In a sense,
this definition of vacuum is for a Wick rotated model,
since the definition is the same as the one in Euclidean orbifolds,
such as, in the ${\mathbb C}/{\mathbb Z}_N$ orbifold.
On the other hand, the right hand side may be interpreted as the
contribution of zero modes, where the vacuum state
is defined by \eqref{vacuum}. 
Thus, it is natural to regard $\rho (\omega^2)$ as the density of
state in the Lorentzian model itself (not in the Wick rotated one).

\section{D-branes and annulus amplitudes}
\label{Dbrane}

The main purpose of this paper is to investigate how D-branes
behave in time-dependent backgrounds, particularly in the
presence of big crunch/big bang singularity.
Since we have D0-branes and D1-branes in the covering space,
we can construct D0-branes and D1-branes even in the orbifold model
as seen in the next subsection.
In subsection \ref{localizedD}, we compute one-loop amplitude
for open strings on the D0-brane, and identify that the 
imaginary part of the amplitude as the pair creation rate
of open strings. Then, we examine D1-brane in subsection \ref{wrappingD},
where annulus amplitude is computed and the imaginary part is identified
as the emission rate of closed strings.

\subsection{Classical analysis on D-branes}
\label{DBI}

In the two dimensional Minkowski space-time ${\mathbb R^{1,1}}$,
we have D0-brane passing from infinite past
to infinite future, and D1-brane with or without constant gauge flux
on its world-volume.
Since the Misner space is regarded as a orbifold of 
${\mathbb R}^{1,1}$, D-branes in Misner space may be obtained by 
utilizing the orbifold method. 
We will also find the D0-brane trajectory and the consistent gauge 
flux on D1-brane by directly solving the DBI actions on the
D-branes as in \cite{HNP}.

\begin{figure}
\centerline{\scalebox{0.6}{\includegraphics{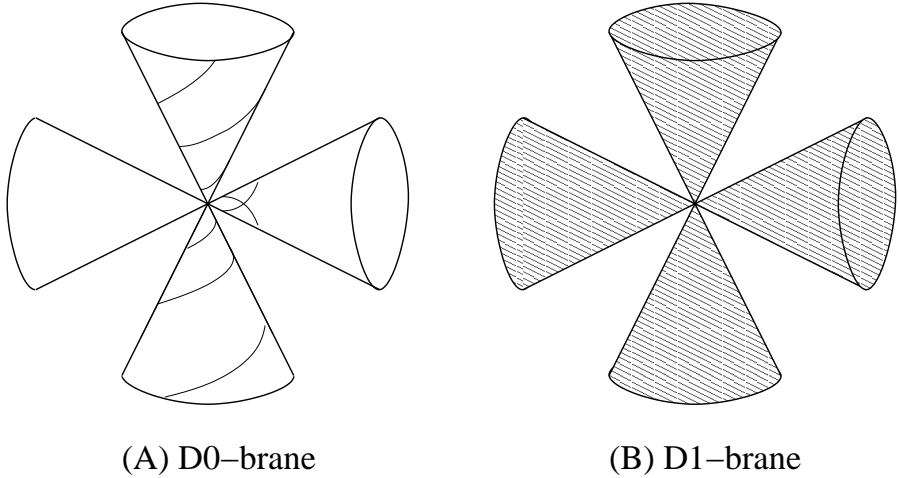}}}
\caption{\it (A) D0-brane travels from infinite past to infinite future
by passing a whisker region. (B) D1-brane covers the whole Misner space
with or without flux on it.}
\label{D0D1misner}
\end{figure}

D0-brane in the covering space is given by a straight line
\begin{align}
 x^{\pm} = x_0^{\pm} + p^{\pm} u
\end{align}
with a parameter $u \in {\mathbb R}$.
Using the coordinate transformations given above and removing
the parameter $u$, this line can be expressed as
\begin{align}
t \sinh (\gamma \psi + A) &= C ,
&e^A&=\sqrt{\frac{p^-}{p^+}},
&C&=\sqrt{\frac{p^+ p^-}{2}}
 \left(\frac{x^+_0}{p^+} - \frac{x^-_0}{p^-} \right)
 \label{D0cline}
\end{align}
for the cosmological regions, and
\begin{align}
r \cosh (\gamma \tau + A) &= C
\label{D0wline}
\end{align}
with the same $A,C$ for the whisker regions.
In the orbifold model, we have to sum up all trajectories with
$A \to A + 2 \pi \gamma k$ with $k \in {\mathbb Z}$. 
Thus, the D0-brane starts from far past in the cosmological
region $t \sim -\infty$ and approaches spirally into the big crunch
singularity. Then it crosses a whisker region and goes to the
other cosmological region (see fig.~\ref{D0D1misner}).

The DBI action for effective theory on the D0-brane in 
the cosmological regions is
\begin{align}
 S &= - \tau_0 \int dt \sqrt{1 - \gamma^2 t^2 \dot \psi ^2} ~,
&\dot \psi &= \frac{d \psi}{d t} ~. 
\end{align}
The normalization includes the inverse of the string coupling
$\tau_0 \propto g_s^{-1}$. 
{}From the equation of motion, we obtain
\begin{align}
 P_{\psi} &= \frac{\delta {\cal L}}{\delta \dot \psi}
          = \frac{\tau_0 \gamma ^2 t^2 \dot \psi}
                 {\sqrt{1 - \gamma ^2 t^2 \dot \psi}} ~,
&\dot \psi ^2 = \frac{1}
 {\gamma ^2 t^2 (1 + \frac{\tau_0^2}{P^2_{\psi}}\gamma ^2 t^2 )} ~,
\end{align}
which is consistent with \eqref{D0cline} if
$C^2 = \frac{P^2_{\psi}}{\gamma ^2 \tau_0^2}$.
The DBI action in the whisker regions is
\begin{align}
S &= - \tau_0 \int d \chi \sqrt{\gamma^2 r^2 - \dot r^2} ~,
&\dot r &= \frac{d r}{d \chi} ~.
\end{align}
Since the energy is conserved, we obtain
\begin{align}
 E &= \frac{\delta {\cal L}}{\delta \dot r} \dot r - {\cal L}
          = \frac{\tau_0 \gamma ^2 r^2}
                 {\sqrt{\gamma^2 r^2 - \dot r^2}} ~,
&\dot r ^2 = \gamma ^2 r^2 - \frac{\tau_0^2}{E^2} \gamma^4 r^4 ~,
\end{align}
which is consistent with \eqref{D0wline} if
$C^2 = \frac{E^2}{\gamma ^2 \tau_0^2}$.

For D1-brane, we can construct D-brane wrapping the whole Misner space,
and add a constant gauge flux on it in the covering space.
In the cosmological regions with the coordinate system 
\eqref{cosmologicalr}, we have the DBI action 
\begin{align}
 S = - \tau_1 \int dt d \psi \sqrt{\gamma ^2 t^2 -F_{t\psi}^2} ~.
\end{align}
Thus the gauge flux can be computed from the Gauss constraint
\begin{align}
 \frac{\delta {\cal L}}{\delta F_{t\psi}} &=
 \frac{\tau_1 F_{t\psi}}{\sqrt{\gamma^2  t^2 - F_{t\psi}^2}} = \Pi ~,
 &F_{t\psi}^2 &= \frac{\gamma ^2 t^2}{1 + \frac{\tau_1^2}{ \Pi^2}} ~.
 \label{gfonD1c}
\end{align}
In the whisker regions, similar analysis leads to the gauge field
\begin{align}
  F_{\chi r}^2 = \frac{\gamma^2 r^2}{1 + \frac{\tau_1^2}{\Pi^2}} ~.
  \label{gfonD1w}
\end{align}
Using the relations
\begin{align}
 F_{t \psi} &= - \gamma t F_{+-} ~,
 &F_{\chi r} & = - \gamma r F_{+-} ~,
\end{align}
we can see that the time-dependent gauge flux \eqref{gfonD1c} 
and \eqref{gfonD1w} in this coordinate system corresponds 
to constant flux in $x^{\pm}$ coordinate system, which is natural 
in the covering space.
Because of this fact, we can solve the open string theory
on D1-brane even with the time-dependent gauge flux using
the orbifold method.

Due to the non-trivial gauge flux on D-branes, open strings
feel the background metric in a modified way \cite{SW}.
We denote the induced closed string metric on D-brane as $g_{ab}$
and the closed string coupling as $g_s$.
Then, open string metric $G_{ab}$
and open string coupling $G_s$ in a configuration with
${\cal F} = B + F$ are given by
\begin{align}
 G_{ab} &= g_{ab} - {\cal F}_{ac} g^{cd} {\cal F}_{db} ~,
 &G_s \equiv e^{\Phi_o}
 &= g_s \sqrt{\frac{- \det G}{ - \det (g + {\cal F})}} ~.
\end{align}
The open string metrics are computed as
\begin{align}
 ds^2 &= \frac{1}
  {1 +\frac{\Pi^2}{\tau_1^2}} (- dt^2 + \gamma^2 t^2 d\psi^2) ~,
&ds^2 &= \frac{1}
  {1+\frac{\Pi^2}{\tau_1^2}} (dr^2 - \gamma^2 r^2 d\chi^2) ~,
\end{align}
and the open string coupling is $G_s = g_s/\sqrt{1+\Pi^2/\tau_1^2}$,
which differ from closed string ones by total normalization.
The low energy spectra for open strings may be read from the
eigenfunctions of Laplacians in terms of open string quantities:
\begin{align}
 \Delta = \frac{1}{e^{- \Phi _o} \sqrt{-G}}
   \partial_a e^{- \Phi _o} \sqrt{-G} G^{ab} \partial_b ~.
   \label{olaplacian}
\end{align}
Since the open strings see the same metric and string coupling
up to normalization, the open string spectrum is the same as the
closed string one. This is consistent with the fact that the
open string spectrum is not changed by introduction of constant
gauge flux in the covering space.

The parameter
\begin{align}
 \Theta^{ij} = - \left(\frac{1}{g+{\cal F}}
    {\cal F} \frac{1}{g-{\cal F}} \right)^{ij} ~,
\end{align}
which represents the non-commutativity of space-time,
can be computed as
\begin{align}
 (\Theta^{t\psi})^2 &= \frac{\tau_1^4}{\gamma ^2 t ^2 \Pi^4}
         \left( 1+\frac{\Pi^2}{\tau_1^2} \right) ~,
 &(\Theta^{\chi r})^2 &=
  \frac{\tau_1^4}{\gamma ^2 r ^2 \Pi^4}
    \left( 1+\frac{\Pi^2}{\tau_1^2} \right) ~.
                      \label{thetap}
\end{align}
Thus we can see that the effective theory has time-dependent
non-commutativity. Notice that this theta parameter is
again constant in the covering space due to the relations
\begin{align}
 \Theta^{t \psi} &= - \frac{1}{\gamma t} \Theta^{+-} ~,
&\Theta^{\chi r} &= - \frac{1}{\gamma r} \Theta^{+-} ~,
\end{align}
and hence we can solve the effective theory by utilizing the
results with a constant non-commutative parameter.
This is quite interesting fact since
there are only few solvable models with time-dependent
non-commutativity.

\subsection{Localized D-brane}
\label{localizedD}

We first deal with D$p$-branes localized in the spatial direction
of the Misner space part. Utilizing the orbifold construction, 
we can construct the D-brane by summing up all image branes
in the covering space. More explicitly, we put a D0-brane
in two dimensional Minkowski space-time at
$e^{2 \pi a} x^+ + b = e^{-2 \pi a} x^- - b$ with
$a,b \in {\mathbb R}$, then the $k$-th image branes are at
$e^{2 \pi (a + \gamma k)} x^+ + b = e^{-2 \pi (a + \gamma k)} x^- - b$.
Speaking differently, we have to consider all open strings 
stretched between these mirror branes.
First we compute annulus amplitude for open strings between these
branes, and then construct boundary state for the D-brane.
We find the imaginary part of the one-loop amplitude, which
implies the open string pair creation between these image branes.

\subsubsection{Annulus amplitude}

Let us compute annulus amplitude for open strings on
a D-brane localized in the spatial direction of Misner space.
Since there may be a subtlety in the definition of spectrum
as mentioned above,
we first derive it in the path integral formalism.
Later we will check that the oscillator formalism gives the same result.
As in the closed string case, we use Euclidean
worldsheet $(\sigma_1 , \sigma_2 )$ with $0 \leq \sigma_1 \leq \pi$
and $\sigma_2 \sim \sigma_2 + 2\pi$, whose metric is
\begin{align}
 ds^2 &= d \sigma_1^2 + t^2 d \sigma_2^2 ~.
\end{align}
The boundary condition in this case is
\begin{align}
 \partial_{\sigma_1} ( e^{2 \pi (a + \gamma k)} X^+ + e^{ - 2 \pi (a + \gamma k) } X^- ) &= 0 ~,
 &\partial_{\sigma_2} ( e^{2 \pi (a + \gamma k)} X^+ - e^{ - 2 \pi (a + \gamma k)} X^- ) &= 0
\label{bcD0pi}
\end{align}
at $\sigma_1 = 0$ and those replaced with $k \leftrightarrow k'$
at $\sigma_1 = \pi$.
Moreover, the boundaries of the worldsheet must be on the branes
\begin{align}
 e^{2 \pi (a + \gamma k)} X^+ + b = e^{ - 2 \pi (a + \gamma k)} X^- - b
\end{align}
at $\sigma = 0$ and those with $k \leftrightarrow k'$
at $\sigma = \pi$.
The mode expansions are then ($\nu = 2 \gamma (k - k')$)%
\footnote{We only consider $\nu \neq 0$ case since the case
with $\nu = 0$ is almost the same as the flat space case.}
\begin{align}
\begin{aligned}
 X^{\pm} (\sigma_1 , \sigma_2 ) &= \mp b +\sum_{m \in {\mathbb Z}}
    e^{\mp 2 \pi (a + \gamma k) \pm \nu \sigma_1+ i m \sigma_2}
     a_{m,0} \\
  &+ \sum_{m \in {\mathbb Z},n>0}
    e^{\mp 2 \pi (a + \gamma k) +i m \sigma_2}
     \left( a^{\pm}_{m,n} e^{ i ( n \mp i \nu) \sigma_1}
     +  a^{\mp}_{m,n} e^{- i ( n \pm i \nu) \sigma_1} \right) ~.
\end{aligned}
\end{align}
Performing the path integral as in the closed string case, we obtain
\begin{align}
 {\cal A}_{k,k'}(t) =
 \left | \prod_m (m - t \nu) \prod_{m,n>0} (m - it(n-i\nu))(m+it(n+i\nu))
 \right |^{-1} =
 \frac{ \eta (it)}{\vartheta_1 (t | \nu| |it)}
   e^{ - \pi t \nu^2} ~.
   \label{D0pfpi}
\end{align}
Notice that the final expression does not depend on the 
parameters $a,b$ of the D0-brane orbit.\footnote{%
It may sound strange that the annulus amplitude does not
depend on the D0-brane orbit since the geodesic
crossing the point $x^+ = x^- = 0$ is incomplete \cite{HE}.
Thus, we may be able to say that the D0-brane does not
feel the real singular point $x^+ = x^- = 0$ as in the case
of Nappi-Witten model \cite{HNP}, though the precise meaning
is obscure.}

Next, we derive the same result in the oscillator formalism.
Because of the boundary condition \eqref{bcD0pi},
the mode expansions are given by
\begin{align}
 X^{\pm} (\tau,\sigma)&= \mp b +
 i \sqrt{\frac{\alpha '}{2}} \sum_n e^{\mp 2 \pi ( a + \gamma k)}
  \left( \frac{\alpha_n^{\pm}}{n \pm i \nu} e^{-i( n \pm i \nu)(\tau + \sigma)}
   + \frac{\alpha_n^{\mp}}{n \mp i \nu} e^{-i( n \mp i \nu)(\tau - \sigma)} \right)~,
\end{align}
where the commutation relations for the oscillators are
$[ \alpha_m^+ , \alpha_n^- ] = ( - m - i \nu ) \delta_{m+n}$.
The Virasoro generator is given by\footnote{As in the closed string case,
we should assign for $\nu < 0$ the creation and annihilation modes 
to the zero modes in the opposite way to $\nu > 0$ case.}
\begin{align}
 L_0 = \frac{1}{2} i \nu (1- i \nu)
 + \sum_{n \geq 1} \alpha^+_{-n} \alpha^-_{n}
 + \sum_{n \geq 0} \alpha^-_{-n} \alpha^+_{n} ~.
\end{align}
Then we find the partition function as
\begin{align}
 {\cal A}_{k,k'} = {\rm Tr}_{{\cal H}_{k,k'}} e^{- 2 \pi t (L_0 - \frac{1}{12})}
 = \frac{ \eta (it)}{\vartheta_1 (t | \nu | |it)}
   e^{ - \pi t \nu^2} ~,
   \label{pfD0}
\end{align}
which reproduces \eqref{D0pfpi}.

Open strings on D$p$-brane have other parts;
ghost part and extra $d$ dimension part with $p$ directions
with Neumann boundary condition and $d-p$ directions
with Dirichlet boundary condition.
Taking these parts into account, the full partition function
is given by
\begin{align}
{\cal A}
 = \sum_{k,k'} \int \frac{dt}{t}
     \frac{ e^{ - \pi t \nu^2}}
  { (8 \pi^2 \alpha ' t)^{\frac{p}{2}}
     \vartheta_1 (t | \nu | |it) \eta (it)^{d-3}} ~.
\label{pfD0f}
\end{align}
This expression is essentially the same as obtained in \cite{Bachas}.
Performing a modular transformation $(t \to s = 1/t)$,
we can rewrite the amplitude in the closed string channel
\begin{align}
{\cal A}
 = \sum_{k,k'} \int ds
   \frac{i}{(8 \pi^2 \alpha ')^{\frac{p}{2}} s^{\frac{d+1-p}{2}} 
     \vartheta_1 ( i | \nu | |i s) \eta (i s)^{d-3}} ~.
 \label{pfD0m}
\end{align}
As seen below, we can reproduce this result as an overlap
between boundary states. 

\subsubsection{Boundary state}

Let us reinterpret the annulus amplitude in terms of closed strings,
in other words, we rewrite the amplitude as an overlap between
boundary states. In the closed string channel, we can see how
closed strings couple to D-branes.
Since the D0-brane is a point-like object in Misner space, 
winding strings do not couple to the D-brane due to its 
macroscopic size.
We first construct boundary state for the original
D0-brane in the covering space, and sum over every boundary
states for image branes mapped by the discrete boost.
We prepare the boundary state for the original D0-brane satisfying
\begin{align}
 (e^{\pm 2 \pi a} \alpha^{\pm}_n +
 e^{\mp 2 \pi a} \tilde \alpha^{\mp}_{-n}) | \tilde D0 \rangle \rangle = 0
\end{align}
for $n \neq 0$ and
\begin{align}
 \sinh (2 \pi a) t + \cosh (2 \pi a) x + \sqrt 2 b &= 0 ~,
&x^{\pm} &= \frac{1}{\sqrt 2} (t \pm x)
\end{align}
for zero modes.
The boundary state can be written explicitly in terms of
oscillators as
\begin{align}
 | \widetilde{D0} \rangle \rangle = \int d t d p
 e^{ \sum_{n \geq 1} \frac{1}{n} ( e^{4 \pi a}
  \alpha^+_{-n} \tilde \alpha^+_{-n}
   + e^{- 4 \pi a}\alpha^-_{-n} \tilde \alpha^-_{-n} ) }
  e^{ip(\sinh (2 \pi a) t + \cosh (2 \pi a) x + \sqrt 2 b)}
| t , p \rangle ~.
\end{align}
Since the orbifold action is generated by\footnote{%
The zero modes satisfy $[x^{\pm} , p^{\mp}] = -i$.}
\begin{align}
g &= \exp ( 2 \pi i \gamma \hat J) ~,
 & i \hat J &= i x^- p^+ - i x^+ p^- 
  + \sum_{n \geq 1} (N_n^+ - N_n^- + \tilde N_n^+ - \tilde N_n^-) ~,
\end{align}
the boundary state is expressed as
\begin{align}
| D0 \rangle \rangle &=
 \sum_{k} g^k
  | \widetilde{D0} \rangle \rangle = \sum_k | D0,k \rangle \rangle ~,
\end{align}
where we define $(\kappa = a + k \gamma)$
\begin{align}
 |D0 , k\rangle \rangle = \int d t d p
 e^{ \sum_{n \geq 1} \frac{1}{n} ( e^{4 \pi \kappa}
  \alpha^+_{-n} \tilde \alpha^+_{-n}
   + e^{- 4 \pi \kappa }\alpha^-_{-n} \tilde \alpha^-_{-n} ) }
  e^{ip(\sinh (2 \pi \kappa ) t
   + \cosh (2 \pi \kappa ) x + \sqrt 2 b)}
| t , p \rangle ~.
\end{align}
This may be also obtained by solving the condition for
the boundary state directly.

Having the boundary state for the D0-brane, we can compute
the overlap between the boundary states from
\begin{align}
 {\cal A}_{k_2,k_1} (s) =
  \langle \langle D0,k_2 | e^{\pi i s (L_0 + \tilde L_0 - \frac{1}{12})}
 | D0,k_1 \rangle \rangle ~.
\end{align}
Non-trivial contribution may arise from zero mode part.
In order to compute the zero mode contribution,
it is convenient to rewrite as \cite{BVC}
\begin{align}
 &\frac{1}{\cosh (2\pi \kappa) }
 \int dt d p  e^{ip(\tanh (2 \pi \kappa) t 
   + x '  )} | t,p \rangle ~,
   &x ' &= x +  \sqrt2 b \, { \rm sech} \, (2 \pi \kappa ) ~.
\end{align}
Then we find 
\begin{align}
\begin{aligned}
 &\frac{  1}{\cosh (2\pi \kappa_1) \cosh (2\pi \kappa_2 ) }
 \int dt dp e^{ip( [\tanh (2 \pi \kappa_1) - \tanh (2 \pi \kappa_2 )]t
     + x ' _1 - x ' _2) - \pi s \alpha ' p^2 /2}
 \\
     &\qquad =  \frac{1}{\cosh (2\pi \kappa_1) \cosh (2\pi \kappa_2)
         |\tanh (2\pi \kappa_1) - \tanh (2\pi \kappa_2)|}
     = \frac{1}{\sinh(\pi | \nu | )} ~.
\end{aligned}
\end{align}
Combining the contribution from the non-zero modes, we obtain
\begin{align}
 {\cal A} (s)
   = \frac{i e^{- \frac{\pi s}{6}}}{\sin (\pi i | \nu | )
      \prod_{n \geq 1} ( 1 - e^{- 2 \pi ( n s -  \nu )})
    ( 1 - e^{- 2 \pi ( n s +  \nu )})} ~.
    \label{D0D0}
\end{align}

The full boundary state for D$p$-brane 
is given by the product
\begin{align}
 | Dp , loc \rangle \rangle =  {\cal N}_{p} \sum_k | B_{gh} \rangle \rangle \otimes
 | D0, k \rangle \rangle \otimes | Bp \rangle \rangle ~.
\end{align}
where $| B_{gh} \rangle \rangle$ denotes the ghost part.
For $d$ free boson part, we have
\begin{align}
 | Bp \rangle \rangle
  = e^{- \sum_{n \geq 1} \frac1n (\alpha^{\beta}_{-n}
  \tilde \alpha^{\beta}_{-n} - \alpha^{I}_{-n}
  \tilde \alpha^{I}_{-n} )} \delta^{(d-p)} ( \hat x - x ) | 0 \rangle ,
  \label{freeboson}
\end{align}
where $2 \leq \beta \leq p+1$ for the directions parallel to
the D$p$-brane and $p+2 \leq I \leq d+1$
for the transverse directions.
The position of the D$p$-brane
in the transverse directions is represented as $x$.
The coefficient is
\begin{align}
{\cal N}_p = 2^{-\frac{d+4}{4}} \sqrt{\pi} (2 \pi \sqrt{\alpha
'})^{\frac{d}{2}-p-1} ~,
\end{align}
which is also used for the boundary state of usual D$p$-brane.
Now the overlap between the boundary states is computed as
\begin{align}
 \langle \langle Dp , loc | \Delta | Dp , loc \rangle \rangle
  = \frac{\alpha '}{2 \pi}
  \int ds  \langle \langle Dp , loc |
  e^{- \pi s (L_0 + \tilde L_0 - \frac{c}{12})}
  | Dp , loc \rangle \rangle ~,
\end{align}
which can be shown to reproduce \eqref{pfD0m}.
Here we used the central charge including the ghost part $c=d-24$.
If we pick up each pair of image branes, the above computation is
exactly the same as in \cite{BVC} as the interaction between
co-moving D$p$-branes.

\subsubsection{Open string pair creation}

We have obtained the one-loop amplitude in the form of integration
over $t$ \eqref{pfD0f}. However, along
the integration contour, the amplitude diverges at
$t=n/ | \nu |$ with $n = 1,2,\cdots$, where $\vartheta_1$ in the
denominator vanishes. In order to avoid the divergence, we shift
the integration contour a little in the complex plane, which
leads to an imaginary part of the partition function.
Picking up the poles, the imaginary part of the amplitude
may be summarized as
\begin{align}
  - 2 {\rm Im}\, {\cal A}
  = \sum_{k,k'}
   \sum_{n=1}^{\infty} \frac{(-1)^{n+1}}{ n (8 \pi ^2 \alpha ')^{\frac{p}{2}}}
    \left( \frac{|\nu|}{n} \right)^{\frac{p}{2}}
     \sum_{states} e^{- 2 \pi n \frac{N-\frac{d}{24}}{|\nu|} - \pi n |\nu|} ~,
\label{paircreation}
\end{align}
where we have defined
$\eta (it)^{-d} \equiv \sum_{states} e^{- 2 \pi t (N - \frac{d}{24})}$.

The situation is very similar to that of \cite{Bachas},
where the annulus amplitude of open strings between co-moving
D-branes is computed. In our case, the co-moving D-branes
are image branes mapped by the discrete boost in the covering space.
The system of co-moving D-branes is T-dual to the configuration
with two D-branes with different electric flux, and
the open strings between these branes are analogue to
charged particles in electric flux. It is well-known that
charged particles are pair created in background with
electric flux by non-perturbative effect \cite{Schwinger}.
It is understood that the pair creation of charged particles 
acts to dilute the background flux.
The pair creation rate may be deduced from the imaginary
part of one-loop amplitude by using the optical theorem.
The open string analogue was given in \cite{BP}, where the
imaginary part of annulus amplitude for open strings
was computed and compared with the charged particle case.
In that case, open strings are pair created due to the
electric flux on D-branes. In the T-dual picture,
the annulus amplitude may be interpreted as the
phase shift due to the scattering between co-moving D-branes.
Therefore, the imaginary part of the amplitude can
be interpreted as the rate that one of the D-branes is
absorbed into another D-brane.
The open string pair creation occurs when the D-branes
are close to each other, and the effect reduces the relative
velocities \cite{Bachas}.
This interpretation 
is T-dual to the fact that the pair creation of charged particles 
dilutes the background flux.

Therefore, also in our case, the imaginary part \eqref{paircreation}
can be interpreted as the absorption rate between image branes
and also the rate of open string pair creation.
Note that the density of states behaves as
$\rho_N \sim e^{2 \pi N}$ for high energy,
thus the imaginary part diverges for $n/|\nu| < 1$.%
\footnote{%
The pair of image branes with $|\nu| > 1$ move faster than 
the speed of light relatively,
which implies the instability of the brane system. This situation is
T-dual to the D-branes with supercritical gauge flux, where the 
open strings are enlarged by the electric force 
and would be of infinite length.}
This means that the probe D-brane localized in the spatial direction of
Misner space is unstable due to the large absorption rate between
mirror branes.
The situation is analogous to the closed string case,
where image probe particles create large black holes due to large
back reaction \cite{HP}.

\begin{figure}
\centerline{\scalebox{0.6}{\includegraphics{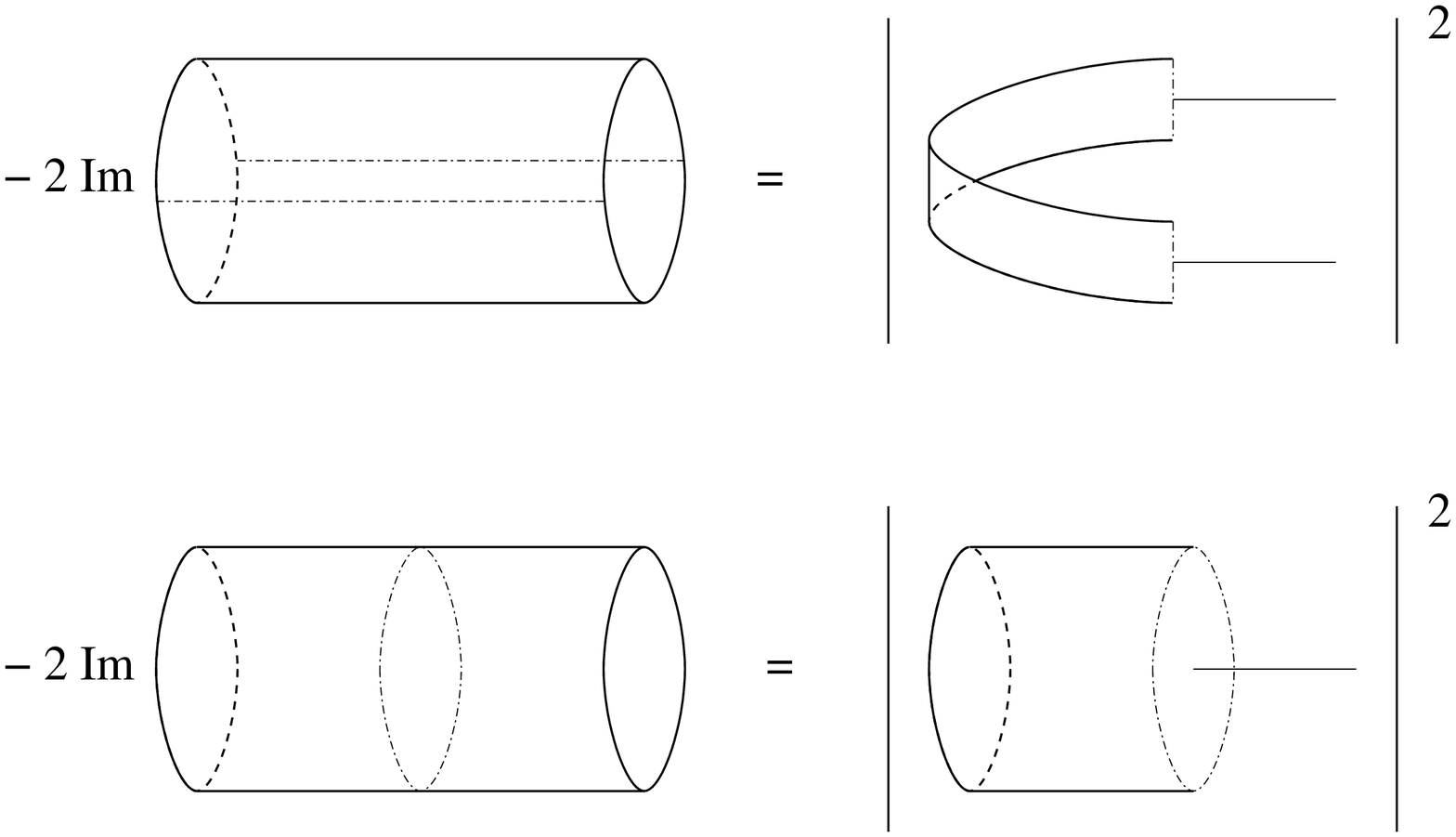}}}
\caption{\it The imaginary part of the cylinder amplitude may
be interpreted as the rate of open string pair creation or/and
closed string emission from the optical theorem.
We identify that the imaginary part for open strings on the D0-brane
is due to open string pair creation, and that for open strings on
the D1-brane is due to emission of winding closed strings.}
\label{optical}
\end{figure}

Utilizing the analysis in \cite{UVfinite},
we can confirm the imaginary part is due to the open string
pair creation, not due to the emission of closed strings
(see fig. \ref{optical}).
With the help of the identity \eqref{identity} with $T=\pi | \nu | t$,
we can rewrite the partition function \eqref{pfD0f} as
\begin{align}
 {\cal A}
  = \sum_{k,k'} \int_0^{\infty} \frac{dt}{t}
       \int_{- \infty}^{\infty} d ( \omega^2 ) \int \frac{d^p k}{(2 \pi )^p}
         \sum_{states} \rho (\omega^2 )
 e^{- 2 \pi t ( \omega ^2 | \nu | + \alpha ' \vec k^2 + N  - \frac{d}{24} + \frac{\nu^2}{2})} ~.
\end{align}
Defining
${\cal A} = \sum_{k,k'}\int \frac{d^p k}{(2 \pi)^p} {\cal A}_{k,k'}(\vec k)$,
we obtain
\begin{align}
 - \frac{1}{2 \pi \alpha '} \frac{d}{d \vec k^2} {\cal A}_{k,k'}(\vec k)
  &= \int_0^{\infty} dt \int_{- \infty}^{\infty}
    d ( \omega^2 )  \sum_{states}  \rho (\omega^2 )
  e^{- 2 \pi t ( \omega ^2 |\nu| + \alpha ' \vec k^2 + N  - \frac{d}{24} + \frac{\nu^2}{2})}
  \\
  &=\int_{- \infty}^{\infty}
    d ( \omega^2 )  \sum_{states}  \rho (\omega^2 )
     \frac{1}{2 \pi ( \omega ^2 |\nu| + \alpha ' \vec k^2 + N  - \frac{d}{24} + \frac{\nu^2}{2})} ~.
 \nonumber
\end{align}
When performing the integration of $\omega^2$,
we encounter poles arising from the propagator
$1/(L_0 -\frac{d}{24})$, which corresponds to on-shell poles of
open strings. These poles lead to imaginary part,
therefore we can conclude that the imaginary part
\eqref{paircreation} is due to the open string pair creation
as discussed in \cite{Bachas}.

To check the consistence
we will reproduce the previous expression
\eqref{paircreation}. For the purpose, it is useful to notice the identity
\begin{align}
 \frac12 ( 1 + \tanh \pi \omega^2 ) =  \sum_{n=1}^{\infty} (-1)^{n+1}
  e^{2 n \pi \omega ^2}
  \label{tanh}
\end{align}
for $\omega^2 < 0$.
Then, by picking up the poles, we obtain
\begin{align}
  - \frac{1}{2 \pi \alpha '} \frac{d}{d \vec k^2} {\rm Im}\, {\cal A}_{k,k'} (\vec k)
   & = \frac{1}{2 |\nu| } \sum_{n=1}^{\infty} \sum_{states}
    (-1)^{n} e^{- \frac{2 \pi n}{|\nu|}(\alpha ' \vec k^2 + N - \frac{d}{24} + \frac{\nu^2}{2})} ~.
\end{align}
Here we have used $\rho (\omega^2 + i n) = \rho (\omega^2)$
with $n \in {\mathbb Z}$.
Integrating out $\vec k^2$ and $\vec k$, we can get the expression \eqref{paircreation}.

One may wonder what would happen if we use \eqref{pfD0m} to evaluate
the imaginary part instead of \eqref{pfD0f}.
In this case, the imaginary part arises at $s = |\nu| /n$, and this
is due to the Regge trajectories in the closed string spectrum.
In other words, the divergence comes from the sum over the states
with integer multiple of spin $2 n$.
Therefore, it is not appropriate to use this channel for
regarding the cause of the imaginary part as the emission 
of on-shell strings.\footnote{Similar discussions were given in 
\cite{Pioline2} for poles in one-loop amplitudes.}

\subsection{Wrapping D-brane}
\label{wrappingD}

Next we consider D$p$-brane wrapping the whole Misner space
including constant gauge field strength $F^{+-}= f$ in the
covering space.
In order to compute annulus amplitude for open strings,
we have to restrict the Hilbert space into subspace invariant
under the discrete boost using the projection operator
$P = \sum_{k} g^k$ where $g X^{\pm} = e^{\pm 2 \pi \gamma} X^{\pm}$. 
In the boundary state formalism, the winding strings can
couple to the D-brane wrapping the whole Misner space. 
Moreover, we will find that the sum of the twisted sector 
corresponds to the sum over all the twist number $k$. 
Recasting the expression in the closed string channel,
we identify that the imaginary part of the cylinder amplitude is
due to the emission of winding closed strings (see fig.
\ref{optical}).

\subsubsection{Annulus amplitude}

We first compute the annulus amplitude in the oscillator
formalism.
Due to the flux $F^{+-}= f$, the boundary condition is
shifted as
\begin{align}
 \partial_{\sigma} X^{\pm} = \pm f \partial_{\tau} X^{\pm}
\end{align}
at the boundary $\sigma = 0, \pi$. Therefore, the mode expansion
is given by
\begin{align}
 X^{\pm} = x^{\pm} - 2 i \alpha '  p^{\pm} \frac{f \sigma + \tau}{\sqrt{1 - f^2}}
 + i \sqrt{ 2 \alpha '} \sum_n \frac{ \alpha^{\pm}_{n}}{n} e^{i n \tau}
  \cos ( n \sigma \mp i {\rm arctanh}\, f ) ~,
\end{align}
where
\begin{align}
 [x^{\pm} , p^{\mp}] &= -i ~,
&[ \alpha^+_m , \alpha^-_n ] &= - m \delta_{m,n} ~.
\end{align}
Notice that the Virasoro generator is the same as in the case
without gauge flux $(f = 0)$;
\begin{align}
 L_0 = - 2 \alpha ' p^+ p^-
 - \sum_{n \geq 1} \alpha^+_{-n} \alpha^-_n
 - \sum_{n \geq 1} \alpha^-_{-n} \alpha^+_n ~.
\end{align}
The orbifold action $g=\exp (2 \pi \gamma i \hat J )$ is
generated by using
\begin{align}
 i \hat J =  i x^- p^+ - i x^+ p^- + \sum_{n \geq 1} (N_n^+ - N_n^-) ~.
\end{align}
Now we can compute the partition function as
\begin{align}
\begin{aligned}
 {\cal A}_k (t) &=
 {\rm Tr} e^{2 \pi i \nu \hat J} e^{- 2 \pi t (L_0 - \frac{1}{12})}
 \\
   &= \frac{e^{- \frac{\pi t}{6}}}{(2i\sin (  \pi i \nu))^2
      \prod_{n \geq 1} ( 1 - e^{- 2 \pi ( n t - \nu )})
    ( 1 - e^{- 2 \pi ( n t + \nu )})} ~,
    \label{D1a}
\end{aligned}
\end{align}
where $\nu = \gamma k$ and we assume that $k \neq 0$. A
non-trivial contribution has arisen from the zero mode sector,
which can be computed as (see, e.g, \cite{BCR})
\begin{align}
 \int d^2 p \langle {\bf p} | g^k q^{\alpha ' {\bf p}^2}
  | {\bf p} \rangle =
 \int d^2 p \delta ^{(2)} ( (1 - g^k) {\bf p}^2 )
   q^{\alpha ' {\bf p}^2} =
 {\det }^{-1} (1- g^k ) = \frac{1}{4 \sinh ^2 \pi \nu }~,
\end{align}
where we use the fact that the eigenvalues of $g^k$ on $p^{\pm}$
are $e^{ \pm 2 \pi \nu}$.

Next we confirm the calculation of annulus amplitude in terms of
path integral, especially the zero mode contribution.
As in the D0-brane case, we use Euclidean worldsheet $(\sigma_1 ,
\sigma_2 )$ with $0 \leq \sigma_1 \leq \pi$ and $\sigma_2 \sim
\sigma_2 + 2\pi$, where the worldsheet metric and the Laplacian
are given by
\begin{align}
 ds^2 &= d^2 \sigma_1 + t^2 d^2 \sigma_2 ~,
&\Delta &= \frac{1}{t^2}(\partial_2^2 + t^2 \partial^2_1) ~.
\end{align}
We assign Neumann boundary condition at the boundaries of worldsheet
as
\begin{align}
 \partial_1 X^{\pm} (\sigma_1=0 , \sigma_2) &=0 ~,
&\partial_1 X^{\pm} (\sigma_1= \pi , \sigma_2) &=0 ~.
 \label{D1bcpi1}
\end{align}
Due to the identification \eqref{misner},
we have the twisted sector along $\sigma_2$ direction as
\begin{align}
 X^{\pm} (\sigma_1 , \sigma_2 + 2 \pi) =
  e^{\pm 2 \pi \gamma k} X^{\pm} (\sigma_1 , \sigma_2) ~.
 \label{D1bcpi2}
\end{align}
The sum over $k$ corresponds to the projection operator
$P=\sum_k g^k$ in the oscillator formalism.
The solutions to the
conditions \eqref{D1bcpi1} and \eqref{D1bcpi2} are given by
\begin{align}
 X^{\pm} (\sigma_1, \sigma_2) =
\sum_{m \in {\mathbb Z},n \geq 0} a^{\pm}_{m,n}
  e^{i (m \mp i \nu) \sigma_2}
  \left( e^{i n \sigma_1} + e^{- i n \sigma_1} \right)  ~.
\end{align}
In order to evaluate the path integral, we have to take
the gauge flux into account. This is included as the
boundary interaction
\begin{align}
 S_B &= \frac{1}{2 \pi \alpha ' \tau_2} \int_{\partial \Sigma}
    A_+ d X^+ ~,
&A_+ &= F_{+-} X^- ~,
&A_- &= 0 ~,
\end{align}
which is cancelled between the contributions at $\sigma_1 = 0$
and $\sigma_2 = \pi$ in the path integral.
The other contributions are summarized as the inverse of
\begin{align}
\begin{aligned}
 {\rm Det} (- \Delta )
 &=  \left|  \prod_m (m - i \nu )^2
 \prod_{m,n > 0} (m+itn-i\nu )(m-itn-i\nu ) \right| \\
 &=(2i\sin (  \pi i \nu ) )^2 \prod_{n \geq 1} e^{2 \pi n t}
    \left( 1 - e^{- 2 \pi ( n t - \nu )} \right)
    \left( 1 - e^{- 2 \pi ( n t + \nu )} \right) \\
 &= 2 \sinh (  \pi |\nu| ) \frac{\vartheta_1 (i |\nu| | it)}{ i \eta(it)} ~,
\end{aligned}
\end{align}
which reproduce the previous result \eqref{D1a}.

To compute the full partition function, we have to
include contributions from the ghosts and from $d$ free bosons.
Since we assign Neumann boundary condition to $p-1$ bosons and
Dirichlet boundary condition to $d_{\perp}\,( \equiv d+1-p)$ bosons, 
the annulus amplitude becomes
\begin{align}
 {\cal A}
  = \sum_{k \neq 0} \int_0^{\infty} \frac{dt}{t}
    \frac{i}
   {2\sinh (  \pi | \nu | ) (8 \pi^2 t \alpha ')^{\frac{p-1}{2}}
    \vartheta_1 (i | \nu | | it)\eta(it)^{d-3}} ~.
\label{D1pf}
\end{align}
This expression may be written
by performing the modular transformation as
\begin{align}
 {\cal A}
  = \sum_{k \neq 0} \int_0^{\infty} ds
    \frac{ e^{- \pi s \nu ^2}}
   { 2 \sinh (  \pi  | \nu |) (8 \pi^2 \alpha ')^{\frac{p-1}{2}}
     s^{\frac{d_{\perp}}{2}}
   \vartheta_1 ( | \nu | s| is)\eta(is)^{d-3}} ~,
   \label{D1mf}
\end{align}
which will be obtained as a scattering amplitude between
boundary states in the next subsection.

\subsubsection{Boundary state}

The D1-brane can wrap the while Misner space, thus the
winding closed strings with macroscopic length can couple to 
the D-brane. This fact implies that the corresponding boundary state
is constructed from closed strings in the twisted sectors
as well as in the untwisted sector.
Since the untwisted sector is the same as in the flat space case,
we concentrate on the twisted sectors. 
First we consider the case without flux,
then the boundary state satisfies the boundary
conditions for $X^{\pm}$ as
\begin{align}
 (\alpha^{\pm}_{n} + \tilde \alpha^{\pm}_{-n}) | D1, w \rangle \rangle = 0
\label{D1bc}
\end{align}
for all $n \in {\mathbb Z}$, where the label $w$ represents the winding number.
In terms of oscillators, the boundary states are expressed as
$(\nu = \gamma w )$
\begin{align}
 | D1, w \rangle \rangle =
 \exp \left( \sum_{n \geq 1}
  \frac{\alpha^+_{-n} \tilde \alpha^-_{-n}}{n + i \nu}
   + \sum_{n \geq 0}
  \frac{\alpha^-_{-n} \tilde \alpha^+_{-n}}{n - i \nu} \right)
  | 0 , w \rangle ~.
  \label{D1bs}
\end{align}
We should also note that the boundary states are invariant under
the discrete boost
\begin{align}
\exp ( 2 \pi i \gamma (J + \tilde J) )| D1, w \rangle \rangle
 = | D1, w \rangle \rangle ~.
\end{align}
The full boundary states are constructed by summing over all the
twisted sectors.
We can introduce gauge flux on the D-brane by inserting
Wilson line into the boundary state as \cite{CLNY}%
\footnote{We can also introduce non-trivial holonomy as
$W = \exp (i \theta w)$ in $w$-th twisted sector,
which also does not change the overlaps for the same D-brane. 
See \cite{Okuyama} for null brane case.}
\begin{align}
 W = {\rm P} \exp \left( - \frac{i}{2 \pi \alpha '} \int A_+ d X^+
  \right) ~.
\end{align}
When the overlap between two boundary states for the same D-branes 
is computed, the Wilson lines are cancelled with each other. 
Therefore, the cylinder amplitude does not change by the introduction
of gauge flux as before.

Using the ghost part $|B_{gh} \rangle \rangle$ and $d$ free boson part
$ | B(p-1) \rangle \rangle$ \eqref{freeboson} with $(p-1)$ Neumann
and $d_{\perp}$ Dirichlet boundary conditions, 
the full boundary state may be written as
\begin{align}
 | Dp , ext \rangle \rangle = \sum_w {\cal N}_w | B_{gh} \rangle \rangle \otimes
 W | D1,w \rangle \rangle \otimes | B(p-1) \rangle \rangle
\end{align}
with a normalization ${\cal N}_w$. Then the overlap between the
boundary states is computed as
\begin{align}
\langle \langle Dp , ext | \Delta 
   | Dp , ext \rangle \rangle  =
\frac{\alpha '}{2 \pi}
  \int ds  \langle \langle Dp , ext |
 e^{- \pi s (L_0 + \tilde L_0 - \frac{c}{24})}
  | Dp , ext \rangle \rangle ~.
\end{align}
Using the overlaps in the Misner space part\footnote{%
As before, we replaced $\nu$ by $|\nu|$ due to the subtlety of
zero modes.} 
\begin{align}
 \langle \langle D1 , w |
         e^{- \pi s (L_0 + \tilde L_0 - \frac{1}{6} )}
         | D1 , w \rangle \rangle  &=
    \frac{i e^{- \pi s ( i | \nu | ( 1 - i | \nu | )) + \frac{\pi s }{6}}}
            {(1 - e^{- 2 \pi i s | \nu | }) 
       \prod_{n \geq 1} (1- e^{- 2 \pi s(n + i |\nu| )})
                        (1-  e^{- 2 \pi s(n - i |\nu| )})} \nonumber \\
          &=
    \frac{\eta (is) e^{-\pi s \nu^2}}{\vartheta_1 (s | \nu | | is)} ~,
  \label{pfD1m}
\end{align}
we can reproduce the open string partition function \eqref{D1mf}
if we set
\begin{align}
{\cal N}_w = \sqrt{\frac{8 \pi^2 \alpha'}{  2 \sinh (  \pi | \nu |) }}{ \cal N}_p ~.
\label{norm}
\end{align}
Notice that the coefficients are different in each twisted sectors.

\subsubsection{Emission of winding closed strings}

As in the annulus amplitude for the localized D-brane,
the amplitude is given by integrating over the moduli parameter
and the integration leads to imaginary part.
For our purpose, it is not useful to adopt the expression
in the open string channel \eqref{D1pf} 
since the divergence originates from the sum over the states
along Regge trajectories as in the previous case.
Therefore, we investigate the imaginary part 
in the closed string channel \eqref{D1mf}.
The integration of $s$ encounters poles at $ s = n / | \nu | $ with 
$n = 1,2, \cdots$, and the contributions from poles are summarized as
\begin{align}
 -2 {\rm Im}\, {\cal A}
  = \sum_{k \neq 0}
   \sum_{n=1}^{\infty} \frac{(-1)^{n+1}}{2 | \nu | \sinh (\pi | \nu |) (8 \pi ^2 \alpha ')^{\frac{p-1}{2}}}
    \left( \frac{|\nu|}{n} \right)^{\frac{d_{\perp}}{2}}
     \sum_{states} e^{- 2 \pi n \frac{N - \frac{d}{24}}{|\nu|} - \pi n |\nu|} ~.
 \label{emission}
\end{align}
Again the imaginary part diverges for $n/|\nu| < 1$
due to the Hagedorn behavior $\rho_N \sim e^{2 \pi N}$.

Following the previous analysis, we identify the origin
of the imaginary part.
Using the identity \eqref{identity} with $T=\pi s | \nu | $, 
we can rewrite the partition function \eqref{D1mf} as
\begin{align}
 {\cal A}
  &= \sum_{k \neq 0} \frac{\alpha '}{2 \pi }\int_0^{\infty} ds
       \int_{- \infty}^{\infty} d ( \omega ^2 )
         \int \frac{d^{d_{\perp}} k_\perp}{(2 \pi )^{d_\perp}}
         \sum_{states} {\cal N}_w^2  \rho (\omega ^2)
 e^{- 2 \pi s ( \omega^2 | \nu |  + \frac{ \alpha '}{2}
    \vec k_{\perp}^2 + N - \frac{d}{24} + \frac{\nu^2}{2})}  \nonumber \\
   &= \sum_{k \neq 0}\frac{\alpha '}{2 \pi}
   \int_{- \infty}^{\infty} d ( \omega ^2 )
         \int \frac{d^{d_{\perp}} k_{\perp}}{(2 \pi )^{d_\perp}}
         \sum_{states}
         \frac{ {\cal N}_w^2  \rho (\omega ^2) }{2 \pi ( \omega ^2 | \nu | + \frac{ \alpha '}{2}
    \vec k_{\perp}^2 + N - \frac{d}{24} + \frac{\nu^2}{2})} ~,
\end{align}
where we denote $\vec k_{\perp}$ as the momenta for directions
transverse to the D-brane.
The imaginary part arises from the poles of 
the closed string propagator $L_0 + \tilde L_0 - \frac{d}{12} = 0$, 
therefore we identify that the imaginary part \eqref{emission} 
is due to the closed string emission (see fig. \ref{optical}).
Note that the closed strings couple to D-branes only if the
modes satisfy the boundary condition \eqref{D1bc}.
In particular, the quasi zero modes satisfy
$\alpha_0^{\pm} + \tilde \alpha_0^{\pm} = 0$, which says that
only strings \eqref{long} localized in the whisker regions are
emitted from the D-branes.\footnote{%
This might be related to the conjecture \cite{Hagedorn}
that the condensation of strings localized in the regions with CTC
protects the chronology, which is supposed to be the string 
version of chronology protection conjecture \cite{Hawking}.}

The emission of closed strings from D-branes was first discussed 
in \cite{LLM,GIR} in the configuration of decaying D-branes
with rolling tachyon \cite{Sen1,Sen2}. In the context, the
D-brane disappears as the tachyon condenses, and the energy is 
taken out by emitted closed strings. 
Similar observation was given in
\cite{NST,Sahakyan,CLS,NPRT,NRS}, where a D-brane is falling
into a stack of NS5-branes \cite{Kutasov}. The D-brane will
form a bound state with NS5-branes, and the closed strings
carry out the generated energy.
The above examples are dynamics in static backgrounds,
but our case is in a time-dependent space-time.
Thus we expect that the winding strings are emitted from D-branes
due to the time-dependent effects. Actually, the time-dependence
enters through the orbifold action \eqref{misner}, therefore
it is natural that the only closed strings in the twisted sectors
feel the time-dependence and are emitted from the D-brane.\footnote{%
As mentioned before, the untwisted sector is the same as the flat
space case, thus there is no closed string emission from the sector.}
We should remark that the emission rate of winding strings 
diverges for $n/|\nu| < 1$. Since the closed strings are emitted,
the background itself may largely change due to the emission.

Finally, we check that the previous expression \eqref{emission}
can be reproduced by collecting poles of propagator as
\begin{align}
 - 2 {\rm Im}\, {\cal A}
 = \sum_{k \neq 0}  \int \frac{d^{d_{\perp}} k_{\perp}}{(2 \pi )^{d_\perp}}
\frac{\alpha '}{2 \pi}
\frac{{\cal N}_w^2 }{| \nu | } 
 \sum_{n=1}^{\infty} \sum_{states} (-1)^{n+1}
  e^{- \frac{2 \pi n}{|\nu|}( \frac{\alpha '}{2} \vec k_{\perp}^2 + N - \frac{d}{24} + \frac{\nu^2}{2})} ~.
\end{align}
Here we have used the identity \eqref{tanh}.


\section{Superstrings and D-branes in Misner space}

The bosonic string theory includes tachyonic states with
no excitation of oscillator modes, therefore the system
is badly unstable.
In order to see that the instability of D-branes is
not related to the tachyonic modes, we
consider type IIA or type IIB superstring theory.
In this section, we study one-loop
amplitude to obtain tachyon free results.
We only adopt the oscillator formalism, but the same results
should be obtained in the path integral formalism.

\subsection{Torus amplitude}

We consider type IIA or type IIB superstrings on the manifold
$ [ {\mathbb R}^{1,1} / \Gamma ] \times {\mathbb R}^{8}$ that is
a product of Misner space and 8 dimensional flat space. 
In the $w$-th twisted sector,
fermions have to obey the conditions
\begin{align}
 \psi^{\pm} (\tau , \sigma + 2 \pi)
&= - (-1)^A e^{\pm 2 \pi \nu} \psi^{\pm} (\tau , \sigma) ~,
&\tilde \psi^{\pm} (\tau , \sigma + 2 \pi)
&= - (-1)^A e^{\pm 2 \pi \nu} \tilde \psi^{\pm} (\tau , \sigma) ~,
\label{NSbc}
\end{align}
where $A=0$ for NS-sector and $A=1$ for R-sector.
The mode expansions can be given then
\begin{align}
 \psi^{\pm} &= \sum_{r}
  b^{\pm}_{-r} e^{-i(r \pm i \nu)(\tau + \sigma)} ~,
 &\tilde \psi^{\pm} &= \sum_{r}
  \tilde b^{\pm}_{-r} e^{-i(r \mp i \nu)(\tau - \sigma)}
\end{align}
with $r \in {\mathbb Z} + (A+1)/2$.
The oscillators satisfy the anti-commutation relations
\begin{align}
 \{ b^{\pm}_r , b^{\mp}_s \} &= - \delta_{r+s} ~,
 &\{ \tilde b^{\pm}_r , \tilde b^{\mp}_s \} &= - \delta_{r+s} ~.
\end{align}
The Virasoro generators are
\begin{align}
\begin{aligned}
 L_0^f &= - \frac{1}{2} \nu^2 - \sum_{r \geq 1/2} (r - i \nu) b^+_{-r} b^-_r
   - \sum_{r \geq 1/2} (r + i \nu) b^-_{-r} b^+_r ~,
  \\
 \tilde L_0^f &= - \frac{1}{2} \nu^2 - 
    \sum_{r \geq 1/2} (r + i \nu) \tilde b^+_{-r} \tilde b^-_r
   - \sum_{r \geq 1/2} (r - i \nu) \tilde b^-_{-r} \tilde b^+_r
\end{aligned}
\end{align}
for NSNS-sector and
\begin{align}
\begin{aligned}
 L_0^f &= \frac{1}{8} - \frac{i \nu ( 1 - i \nu)}{2}
   - \sum_{r \geq 1} (r - i \nu) b^+_{-r} b^-_r
   - \sum_{r \geq 0} (r + i \nu) b^-_{-r} b^+_r ~,
  \\
 \tilde L_0^f &= \frac{1}{8} - \frac{i \nu ( 1 - i \nu)}{2}
   -
    \sum_{r \geq 0} (r + i \nu) \tilde b^+_{-r} \tilde b^-_r
   - \sum_{r \geq 1} (r - i \nu) \tilde b^-_{-r} \tilde b^+_r
\end{aligned}
\end{align}
for RR-sector. The Virasoro generators for NSR and RNS-sectors 
are given in a similar way.

In order to obtain the full partition function, we add flat 8
dimension and ghost part, and assign the GSO projection necessary 
to project out the unphysical states. 
Denoting $F,\tilde F$ as the worldsheet
fermion number operators, the partition function can be computed
as
\begin{align}
\begin{aligned}
 {\cal T} = \int_{\cal F} \frac{d^2 \tau}{\tau_2} \sum_{w,k}
     &\left[  {\rm Tr}^w_{NSNS} \left({\textstyle \frac14}
        (1 + (-1)^F)(1 + (-1)^{\tilde F})
         g^k q^{L_0} \bar q^{\tilde L_0} \right)\right.
    \\
    & - {\rm Tr}^w_{NSR} \left( {\textstyle \frac14}
        (1 + (-1)^F)(1 \mp (-1)^{\tilde F})
         g^k q^{L_0} \bar q^{\tilde L_0} \right)
    \\
   &  -  {\rm Tr}^w_{RNS} \left( {\textstyle \frac14}
        (1 + (-1)^F)(1 + (-1)^{\tilde F})
         g^k q^{L_0} \bar q^{\tilde L_0 } \right)
    \\ &\left.
     +  {\rm Tr}^w_{RR} \left( {\textstyle \frac14}
        (1 + (-1)^F)(1 \mp (-1)^{\tilde F})
         g^k q^{L_0} \bar q^{\tilde L_0} \right) \right] ~,
\end{aligned}
\end{align}
where we use different GSO projections for type IIA and type IIB.
Namely, we assign $-$ for type IIA and $+$ for type IIB.
The trace Tr$^w$ is taken in the $w$-th twisted sector and the
boost operator $g = \exp (2 \pi \gamma i (J + \tilde J))$
is defined by
\begin{align}
 i J &= - \sum_{r \geq \frac12} b^+_{-r} b^-_r 
       +  \sum_{r \geq \frac12} b^-_{-r} b^+_r ~,
 &i \tilde J &= - \sum_{r \geq \frac12} \tilde b^+_{-r} \tilde b^-_r 
       +  \sum_{r \geq \frac12} \tilde b^-_{-r} \tilde b^+_r
       \label{JNSNS}
\end{align}
for NSNS-sector,
\begin{align}
 i J &= - \sum_{n \geq 1} b^+_{-n} b^-_n 
       +  \sum_{n \geq 0} b^-_{-n} b^+_n ~,
 &i \tilde J &= - \sum_{n \geq 0} \tilde b^+_{-n} \tilde b^-_n 
       +  \sum_{n \geq 1} \tilde b^-_{-n} \tilde b^+_n
       \label{JRR}
\end{align}
for RR-sector, and similarly for NSR and RNS-sectors. 
This one-loop amplitude can be evaluated as \cite{Nekrasov}
\begin{align}
 {\cal T} &=
 \int_{\cal F} \frac{d^2 \tau}{16 \pi^2 \tau_2^2 \alpha '}
  \sum_{k,w}
  \frac{|\vartheta_3 (y|\tau) \vartheta_3 (0|\tau)^3
        - \vartheta_4 (y|\tau) \vartheta_4 (0|\tau)^3
        - \vartheta_2 (y|\tau) \vartheta_2 (0|\tau)^3
        |^2}{(4 \pi^2 \tau_2 \alpha ')^3
             |\vartheta_1 (y|\tau ) \eta (\tau )^9|^2}
\end{align}
with $y=i \gamma ( w \tau + k )$.
This can be expressed as
\begin{align}
 {\cal T} =
  \int_{\cal F} \frac{d^2 \tau}{4 \pi^2 \tau_2^2 \alpha '}
  \sum_{k,w}
  \frac{| \vartheta_1 (y/2|\tau)^4 
        |^2}{(4 \pi^2 \tau_2 \alpha ')^3
             |\vartheta_1 (y|\tau ) \eta (\tau )^9|^2} 
\end{align}
using the identity
\begin{align}
\vartheta_3 (y|\tau) \vartheta_3 (0|\tau)^3
        - \vartheta_4 (y|\tau) \vartheta_4 (0|\tau)^3
        - \vartheta_2 (y|\tau) \vartheta_2 (0|\tau)^3
        = 2 \vartheta_1 (y/2|\tau)^4 ~.
\label{Riemann}
\end{align}
The one-loop amplitude does not vanish because the
supersymmetry is broken though the boundary conditions 
\eqref{twist} and \eqref{NSbc}.
Notice that tachyonic state in NSNS-sector, which is apparent
signal of background instability, is projected out
by the GSO projection.

Before moving to the open string sector, we should comment on
the case with anti-periodic boundary condition for
fermions. Namely, we can consistently assign
\begin{align}
\begin{aligned}
 \psi^{\pm} (\tau , \sigma + 2 \pi)
&= - (-1)^{A + w} e^{\pm 2 \pi \nu} \psi^{\pm} (\tau , \sigma) ~, 
\\
\tilde \psi^{\pm} (\tau , \sigma + 2 \pi)
&= - (-1)^{A + w} e^{\pm 2 \pi \nu} \tilde \psi^{\pm} (\tau , \sigma) ~.
\end{aligned}
\end{align}
The action of GSO projection is not changed for even $w$ but
opposite for odd $w$. Therefore, the tachyonic modes appear
in the odd $w$-th twisted sector, which will be a signal of 
background instability. 
The condensation of twisted closed tachyons
is discussed in \cite{Hagedorn,MS} using the similarity to 
the condensation of localized tachyon in 
Euclidean orbifolds ${\mathbb C}/{\mathbb Z}_N$ \cite{APS}.
Thus, the case with anti-periodic fermions
would be important, if we consider the condensation of winding 
closed tachyons.

\subsection{Localized D-brane}

We first consider D$p$-brane localized in the Misner space part.%
\footnote{We set $p$ even for type IIA and odd for type IIB as usual.}
More precisely, we consider the D-brane at
$e^{2 \pi a} x^+ + b = e^{- 2 \pi a} x^- - b$
and their images of orbifold action in the covering space.
The boundary conditions
for fermions in Misner space are given by
\begin{align}
\begin{aligned}
  \psi^{\pm} (\tau , \sigma = 0) &=
  e^{\mp 4 \pi (a + \gamma k_1)} \tilde \psi^{\mp} (\tau , \sigma =0) ~,
   \\
  \psi^{\pm} (\tau , \sigma = \pi )
  & = - (-1)^A  e^{\mp 4 \pi (a + \gamma k_2)} \tilde \psi^{\mp} (\tau , \sigma = \pi ) ~,
\end{aligned}
\end{align}
where $A=0$ for NS sector and $A=1$ for R-sector.
Then, the mode expansions read
\begin{align}
 \psi^{\pm} &= \sum_{r}
  b^{\pm}_{-r} e^{\mp 2 \pi (a + \gamma k_1) -i(r \pm i \nu)(\tau + \sigma)} ~,
 &\tilde \psi^{\pm} &= \sum_{r}
  b^{\mp}_{-r} e^{\mp 2 \pi (a + \gamma k_1) -i(r \mp i \nu)(\tau - \sigma)} ~,
\end{align}
with
$ \{ b^{+}_r , b^{-}_s \} = - \delta_{r+s} $.
Here $\nu = 2 \gamma (k_1 - k_2)$ and $ r \in {\mathbb Z} + (A+1)/2$. 
The Virasoro generators are
\begin{align}
 L_0^f &= -\frac12 \nu^ 2 - \sum_{r \geq 1/2} (r - i \nu) b^+_{-r} b^-_r
   - \sum_{r \geq 1/2} (r + i \nu) b^-_{-r} b^+_r
\end{align}
for NS-sector and
\begin{align}
 L_0^f &= \frac{1}{8} - \frac{i \nu ( 1 - i \nu)}{2}
   -  \sum_{r \geq 1} (r - i \nu) b^+_{-r} b^-_r
   - \sum_{r \geq 0} (r + i \nu) b^-_{-r} b^+_r
\end{align}
for R-sector.
For other 8 coordinates, we again assign the Neumann boundary condition
for $p$ coordinates and the Dirichlet boundary condition for $8-p$
coordinates.
Then, the partition function is\footnote{Here we replaced
$\nu$ by $|\nu|$ from the same reason as before.}
\begin{align}
{\cal A} = \int_0^{\infty} \frac{dt}{2t}
\frac{\vartheta_3 ( |\nu| t| it) \vartheta_3 (0|it)^3
             - \vartheta_4 (|\nu| t|it) \vartheta_4 (0|it)^3
             - \vartheta_2 (|\nu| t|it) \vartheta_2 ( 0 |it)^3}
             {(8\pi^2 t \alpha')^{\frac{p}{2}}
              \vartheta_1(|\nu| t|it) \eta(it)^9} ~,
              \label{superD0o}
\end{align}
or by modular transformation $(s = 1/t)$
\begin{align}
{\cal A} = \int_0^{\infty} \frac{i ds}{2}
\frac{\vartheta_3 (i |\nu| | it) \vartheta_3 (0|is)^3
             - \vartheta_4 (i|\nu ||is) \vartheta_4 (0|is)^3
             - \vartheta_2 (i|\nu ||is) \vartheta_2 ( 0 |is)^3}
             {(8\pi^2 \alpha')^{\frac{p}{2}} s^{\frac{8-p}{2}}
              \vartheta_1(i |\nu ||is) \eta(is)^9} ~.
              \label{superD0}
\end{align}
This partition function should be reproduced as a scattering
amplitude between boundary states.

As in the bosonic case, we first construct boundary state
for the original D-brane and generate 
boundary states for the image branes by utilizing the orbifold action.
For the D$p$-brane at $e^{2 \pi a} x^+ + b = e^{- 2 \pi a} x^- - b$,
the boundary state $| Dp,\eta \rangle \rangle_P$
 ($P=NSNS$ or $RR$) satisfies the conditions
\begin{align}
 &( e^{\pm 2 \pi a} b^{\pm}_{-r} + i \eta e^{\mp 2 \pi a}
  \tilde b^{\pm}_r )
 | Dp,\eta \rangle \rangle_P = 0~, \nonumber \\
 &( b^{\alpha}_{-r} + i \eta \tilde b^{\alpha}_r )
 | Dp,\eta \rangle \rangle_P = 0~, \\
 &( b^{I}_{-r} - i \eta \tilde b^{I}_r )
 | Dp,\eta \rangle \rangle_P  = 0~. \nonumber
\end{align}
Here we used indices $2 \leq \alpha \leq p$ for directions
parallel to D$p$-brane and $p +1 \leq I \leq d+1$ for
transverse directions. We also use
$r \in {\mathbb Z} + \frac12$ for $P=NSNS$ and
$r \in {\mathbb Z}$ for $P=RR$, and $\eta = \pm 1$
representing the spin structure.
The image branes are generated as
\begin{align}
 | Dp,k,\eta \rangle \rangle_P = e^{2 \pi i \gamma k (J + \tilde J)}
 | Dp,\eta \rangle \rangle_P ~,
\end{align}
where the boost operator is defined as \eqref{JNSNS} and \eqref{JRR}.
Then the fermionic parts of the boundary state are explicitly written as
\begin{align}
\begin{aligned}
|Dp, k,\eta \rangle \rangle_{NSNS}
 &= e^{ i \eta \sum_{r > 0}
    \left( e^{4 \pi \kappa} b^+_{-r} \tilde b^+_{-r}
      + e^{- 4 \pi \kappa} b^-_{-r} \tilde b^-_{-r}
    - b^{\alpha}_{-r} \tilde b^{\alpha}_{- r} +
      b^I_{-r} \tilde b^I_{-r} \right) } | 0 \rangle  ~,
      \\
|Dp,k,\eta \rangle \rangle_{RR}
 &= e^{ i \eta \sum_{n > 0}  \left(
 e^{4 \pi \kappa} b^+_{-n} \tilde b^+_{-n}
      +  e^{-4 \pi \kappa}b^-_{-n} \tilde b^-_{- n}
- b^{\alpha}_{-n} \tilde b^{\alpha}_{-n}
      + b^I_{-n} \tilde b^I_{- n} \right) }
   | k, \eta \rangle^{(0)}_{RR} 
\end{aligned}
\end{align}
with $\kappa = a + \gamma k$. 
We use the normalization of RR vacuum states as%
\footnote{See, e.g., \cite{DbranereviewI,DbranereviewII} for the 
explicit form and other properties of the RR vacuum states.}
\begin{align}
  | k, \eta \rangle^{(0)}_{RR}  &
   = e^{2 \pi i \gamma k ( J + \tilde J)} | \eta \rangle^{(0)}_{RR} ~,
&{}^{\,(0)}_{RR} \langle \eta_1 | \eta_2 \rangle^{(0)}_{RR}
&= - 16 \delta_{\eta_1 \eta_2} ~.
\end{align}
The boundary state also includes ghost part and bosonic part
with $d=8$ given above.
The GSO invariant combination is
\begin{align}
\begin{aligned}
 | Dp , loc \rangle \rangle &=
  \frac12 \left( | Dp, loc, + \rangle \rangle_{NSNS}
    -  | Dp, loc, - \rangle \rangle_{NSNS} \right) \\
 &+ \frac12 \left( | Dp, loc, + \rangle \rangle_{RR}
    +  | Dp, loc, - \rangle \rangle_{RR} \right) ~,
\end{aligned}
\end{align}
where the sign in front of the RR sector is changed
from $+$ to $-$ for anti-D-brane.

We would like to show the scattering amplitude between the boundary
states reproduces the amplitude \eqref{superD0}.
A non-trivial contribution comes from the zero mode part of RR-sector.
For the part, the boost operator is written as
\begin{align}
 i J + i \tilde J =  b_0^- b_0^+ - \tilde b_0^+ \tilde b_0^-
        = d^t_+ d^x_- + d^t_- d^x_+ ~,
\end{align}
where we denote
\begin{align}
  d^t_{\pm} &= \frac12 [( b_0^+ + b^-_0) \pm i (\tilde b_0^+ + \tilde b^-_0)] ~,
 &d^x_{\pm} &= \frac12 [( b_0^+ - b^-_0) \pm i (\tilde b_0^+ - \tilde b^-_0) ] ~.
\end{align}
Note that the new operators have the following properties as
\begin{align}
 \{ d^i_+ ,d^j_- \} &= \delta^{ij} ~,
&d^t_{\eta} |\eta \rangle_{RR}^{(0)} &= 0 ~,
&d^x_{- \eta} |\eta \rangle_{RR}^{(0)} &= 0 ~.
\end{align}
Then, we obtain by following \cite{BVC} as
\begin{align}
 | k ,\eta >_{RR}^{(0)}
  = e^{2 \pi i \gamma k ( J + \tilde J)} |\eta \rangle_{RR}^{(0)}
  = [ \cosh (2 \pi \gamma k)
  + \sinh  (2 \pi \gamma k)  d^t_{-\eta} d^x_{\eta} ] |\eta \rangle_{RR}^{(0)}
\end{align}
and hence
\begin{align}
  {}^{\,(0)}_{RR} \langle k_1 , \eta | k_2 ,\eta \rangle_{RR}^{(0)}
   = -16 \cosh (2 \pi \gamma | k_1 - k_2 |) ~.
\end{align}
Combining this result and contributions from other fermionic modes, 
bosonic modes and ghost modes, we can show that the overlap reproduces
the partition function \eqref{superD0}.

Finally let us evaluate the imaginary part of the one-loop amplitude
\eqref{superD0o}. As in the bosonic case, the poles in the integration over 
$t$ leads to the imaginary part as
\begin{align}
 - 2{\rm Im}\, {\cal A}
  = \sum_{k,k'}
   \sum_{n=1}^{\infty} \sum_{states}
\frac{(-1)^{(n+1)(A+1)}}{| \nu | (8 \pi ^2 \alpha ')^{\frac{p}{2}}}
    \left( \frac{|\nu|}{n} \right)^{\frac{p}{2}+1}
      e^{- 2 \pi n \frac{N}{|\nu|}} ~,
\label{superpaircreation}
\end{align}
where $A=0$ for bosons (NS-sector) and $A=1$ for fermions (R-sector).
Thus, in the supersymmetric case, the sign depends on whether 
the state is bosonic or fermionic, and the zero point energy shift 
by $\frac12 |\nu|^2$ disappears.
Notice that tachyonic state is projected out from the spectrum
of pair created open string $N$, which includes fermionic sector as 
well. Therefore, we can say that the instability due to the 
open string pair creation is not related to the tachyonic mode.

\subsection{Wrapping D-brane}

Next let us generalize D$p$-brane wrapping the whole Misner space
to superstring case. Just like the bosonic case, we introduce
a constant gauge flux $F^{+-} = f$ on D-brane in the covering space.
We adopt for Misner space part fermions with boundary 
conditions
\begin{align}
\begin{aligned}
 \psi^{\pm} - \tilde \psi^{\pm} 
 &= \pm f ( \psi^{\pm} + \tilde \psi^{\pm} ) &\text{ at } \sigma &= 0 ~,
 \\
 \psi^{\pm} - \tilde \psi^{\pm} 
 &= \mp (-1)^A  f ( \psi^{\pm} + \tilde \psi^{\pm} ) &\text{ at } \sigma &= \pi ~,
\end{aligned}
\end{align}
where $A=0$ and $A=1$ correspond to NS-sector and R-sector, respectively.
Therefore, we have the mode expansions 
\begin{align}
 \psi^{\pm} &= \sum_{r}
  b^{\pm}_{-r} e^{-i r (\tau + \sigma) \pm {\rm arctanh}\, f} ~,
 &\tilde \psi^{\pm} &= \sum_{r}
  b^{\pm}_{-r} e^{-i r (\tau - \sigma) \mp {\rm arctanh}\, f} ~,
\end{align}
with $r \in {\mathbb Z} +(A+1)/2$ and the Virasoro generator
\begin{align}
 L_0^f &= c(A) - \sum_{r > 0} r b^+_{-r} b^-_r
   - \sum_{r > 0} r b^-_{-r} b^+_r
\end{align}
with $c(0)= 0 $ and $c(1)= \frac18$.
Here the oscillators satisfy 
$\{ b^{\pm}_r , b^{\mp}_s \} = - \delta_{r+s}$.
We will use the projection operator defined as
\begin{align}
 P &= \sum_{k} e^{2 \pi i \gamma k \hat J} ~,
 &e^{2 \pi i \gamma \hat J} \psi^{\pm} &= e^{\pm 2 \pi \gamma} \psi^{\pm} ~.
\end{align}
For other 8 coordinates, we assign the Neumann boundary condition for
$p-1$ coordinates and the Dirichlet boundary condition for $9-p$
coordinates.
Taking into account the GSO projection, the partition function
can be written as
\begin{align}
{\cal A}(t) = {\rm Tr}_{{\cal H}_{\rm NS}} P {\textstyle \frac12} (1+(-1)^F)
  e^{-2\pi t L_0 } -
  {\rm Tr}_{{\cal H}_{\rm R}} P {\textstyle \frac12} (1 \pm (-1)^F)
  e^{-2\pi t L_0} ~.
\end{align}
Including ghost parts, we obtain
\begin{align}
{\cal A} = \int_0^{\infty} \frac{idt}{2t}
 \frac{\vartheta_4 (i | k |\gamma|it) \vartheta_4 (0|it)^3
             - \vartheta_3(i |k |\gamma|it) \vartheta_3(0|it)^3
             - \vartheta_2 (i |k |\gamma|it) \vartheta_2 ( 0 |it)^3}
             { 2 \sinh (\pi |k| \gamma )(8\pi^2 t \alpha')^{\frac{p-1}{2}} 
              \vartheta_1(i |k |\gamma|it) \eta(it)^9} ~.
\end{align}
Using the modular transformation, we could write as
\begin{align}
{\cal A} = \int_0^{\infty} \frac{ds}{2}
 \frac{\vartheta_3 ( | k | \gamma s|is) \vartheta_3 (0|is)^3
             - \vartheta_4 ( | k | \gamma s|is) \vartheta_4 (0|is)^3
             - \vartheta_2 ( | k | \gamma s|is) \vartheta_2 ( 0 |is)^3}
             {2 \sinh (\pi |k| \gamma ) (8\pi^2 \alpha')^{\frac{p-1}{2}}s^{\frac{9-p}{2}}
               \vartheta_1( | k | \gamma s|is) \eta(is)^9} ~,
               \label{superD1}
\end{align}
which should be reproduced by an overlap between boundary states.

As in the bosonic case, we construct boundary states only in the
twisted sectors.
The boundary states can be written as the product
\begin{align}
 | Dp, ext, \eta \rangle \rangle_P
  = \sum_w {\cal N}_w | B_{gh.}, \eta \rangle \rangle_P
   \otimes W | D1 , w , \eta \rangle \rangle_P
   \otimes | B(p-1) , \eta \rangle \rangle_P ~,
\end{align}
where $P=NSNS$ or $RR$ and ${\cal N}_w$ is \eqref{norm} with $d=8$.
The Wilson line also includes fermionic part as
\begin{align}
 W &= {\rm P} \exp \left[ - \frac{i}{2 \pi \alpha '} \int_0^{2 \pi} d \sigma \left(
    A_{\mu} \partial_{\sigma} X^{\mu}
    - \frac{i}{2} F_{\mu \nu} \theta^{\mu} \theta^{\nu}
     \right) \right] ~,
     & \theta^{\mu} &= \psi^{\mu} - i \eta \tilde \psi^{\mu} ~,
\end{align}
though it gives no contribution to the overlaps for the same D-brane.
The boundary states given above satisfy the boundary conditions
\begin{align}
\begin{aligned}
 &( b^{\pm}_{-r} + i \eta \tilde b^{\pm}_r )
 |D1, w,\eta \rangle \rangle_P = 0~, \\
 &( b^{\alpha}_{-r} + i \eta \tilde b^{\alpha}_r )
 |B(p-1),\eta \rangle \rangle_P = 0~, \\
 &( b^{I}_{-r} - i \eta \tilde b^{I}_r )
 |B(p-1), w,\eta \rangle \rangle_P = 0~,
\end{aligned}
\end{align}
where $2 \leq \alpha \leq p-1$ for directions tangent to D-brane and
$p \leq I \leq d+1$ for normal directions. 
Picking up the fermionic part, they are expressed as
\begin{align}
\begin{aligned}
|D1, w,\eta \rangle \rangle_{NSNS}
 &= e^{ i \eta \sum_{n \geq 1} \left(b^+_{-n + 1/2} \tilde b^-_{- n + 1/2}
      + b^-_{-n + 1/2} \tilde b^+_{- n + 1/2} \right) } | 0 \rangle  ~,
\\
|D1,w,\eta \rangle \rangle_{RR}
 &= e^{ i \eta \left(\sum_{n \geq 1} b^+_{-n} \tilde b^-_{-n}
      + \sum_{n \geq 0} b^-_{-n} \tilde b^+_{- n} \right) } | 0 \rangle
\end{aligned}
\end{align}
for Misner part and
\begin{align}
\begin{aligned}
|B(p-1),\eta \rangle \rangle_{NSNS}
 &= e^{ -i \eta \sum_{n \geq 1} \left(b^{\alpha}_{-n + 1/2} \tilde b^{\alpha}_{- n + 1/2}
      - b^I_{-n + 1/2} \tilde b^I_{- n + 1/2} \right) } | 0 \rangle  ~,
\\
|D1,w,\eta \rangle \rangle_{RR}
 &= e^{ -i \eta \left(\sum_{n \geq 1} b^{\alpha}_{-n} \tilde b^{\alpha}_{-n}
      - \sum_{n \geq 1} b^I_{-n} \tilde b^I_{- n} \right) }
   | \eta \rangle^{(0)}_{RR}
\end{aligned}
\end{align}
for the other 8 dimension.
The normalization of RR vacuum states is set as
${}^{\,(0)}_{RR} \langle \eta_1 | \eta_2 \rangle^{(0)}_{RR}
= - 8 \delta_{\eta_1 \eta_2}$.
The GSO invariant combination is given by
\begin{align}
\begin{aligned}
 | Dp , ext \rangle \rangle &=
  \frac12 \left( | Dp, ext, + \rangle \rangle_{NSNS}
    -  | Dp, ext, - \rangle \rangle_{NSNS} \right) \\
 &+ \frac12 \left( | Dp, ext, + \rangle \rangle_{RR}
    +  | Dp, ext, - \rangle \rangle_{RR} \right)~.
\end{aligned}
\end{align}
Computing the overlaps between the boundary states,
we reproduce \eqref{superD1}.

The imaginary part of the cylinder amplitude arises 
when integrating \eqref{superD1} by the
moduli parameter $s$.
In order to avoid divergence, we shift the contour and then
pick up the poles as
\begin{align}
 - 2 {\rm Im}\, {\cal A}
  = \sum_{w \neq 0}
   \sum_{n=1}^{\infty}  \sum_{states} 
 \frac{(-1)^{(n+1)(A+1)}}{2 |\nu| \sinh (\pi |\nu|)  (8 \pi ^2 \alpha ')^{\frac{p-1}{2}}}
    \left( \frac{|\nu|}{n} \right)^{\frac{9-p}{2}}
     e^{- 2 \pi n \frac{N}{|\nu|}} ~,
 \label{superemission}
\end{align}
which corresponds to the rate of winding string emission
as discussed in the bosonic case. We can see that
there are no emission of tachyonic modes and 
no zero point energy shift.


\section{Correlation functions}
\label{4pt}

As shown in \cite{LMS1,LMS2,BCKR},
the closed string $2 \to 2$ amplitude in the untwisted sector
diverges due to the graviton exchange near the big crunch/big bang
singularity, which was interpreted as the signal of large back reaction.
Since there is no gravity mode in open string spectrum,
it is natural to expect that the
divergence becomes milder for open string $2 \to 2$ scattering on
wrapping D-brane, which will be examined in this section. The
gauge flux on D-branes induces the (time-dependent) non-commutativity, 
so it is quite interesting to see if the singular behavior is improved
by the non-commutativity.
For definiteness, we consider D$p$-brane wrapping the Misner space
part in the critical bosonic string theory. Moreover, we
introduce gauge flux only on the Misner space part.
The generalization to the superstring case may be done
by following \cite{Pioline3}.

\subsection{Wave function and two point function}

Wave function can be realized as a superposition of plane waves in
the covering space as \cite{Nekrasov,BCKR}\footnote{%
In this section, we set $\gamma = 1$ for simplicity.}
\begin{align}
 \Psi_{m,l} =\sqrt{ \frac{m}{2 \pi}}
  \int d w e^{\frac{i m }{\sqrt2}
   (X^- e^{- w} + X^+ e^{w}) + i w l + i \vec k \cdot \vec X}
\label{wavefunction}
\end{align}
with $m \geq 0,l \in {\mathbb Z}$. We can easily see that the wave
function is invariant under the discrete boost \eqref{misner}.
Using the fact that the shift $p^{\pm} \to e^{\pm a} p^{\pm}$
gives only a phase factor, the momenta satisfying the on-shell
condition $2 p^+ p^- = m^2$ have been set as $p^{\pm}= m / \sqrt2$. This
wave function represents an excitation over the adiabatic vacuum
\cite{BD}, which is defined utilizing the vacuum in the Minkowski
space. This means that the wave function with $p^{\pm} > 0$
represents in-coming wave. We have denoted $\vec k$ as momenta for the
other $(p-1)$ directions parallel to the D$p$-branes.

As mentioned above, we would like to introduce gauge flux on
the D-brane. In the covering space, we condense constant
gauge field strength, which induces non-commutativity \cite{SW}
\begin{align}
 [ x^+ , x^- ] = i \Theta^{+ -} \,(\equiv i \theta) ~.
\end{align}
In the presence of non-zero $\Theta^{ij}$, the correlation functions
for plane waves include non-trivial phase factor as \cite{SW}
\begin{align}
 \left\langle \prod_n e^{i k^n_i X^i }(\tau _n) \right\rangle
   = e^{-\frac{i}{2}\sum_{n > m} k^n_i \Theta^{ij} k^m_j
        \epsilon (\tau_n - \tau_m) } \delta(\sum_n k^n) ~,
        \label{NCcorrelator}
\end{align}
where $\epsilon (\tau_n - \tau_m ) = 1$ for $\tau_n > \tau_m $
and $\epsilon (\tau_n - \tau_m ) = - 1$ for $\tau_n < \tau_m $.
Using this formula, we can compute two point functions of
\eqref{wavefunction} as $(\tau_1 > \tau_2)$
\begin{align}
 \langle \Psi^*_{m_2,l_2} \Psi_{m_1,l_1} \rangle
  &= \frac{2 \sqrt{m_1 m_2}}{\pi} \int dw_+
 e^{i \left( l_1 - l_2  \right) w_+ }
   \int d w_- e^{ i \left( l_1 + l_2 \right) w_-}
    e^{\frac{i}{4}\theta m_1 m_2 (e^{2 w_-} - e^{-2 w_-})}
   \nonumber \\ &\times \delta (m_1 e^{-w_-} - m_2 e^{w_-} )
    \delta (m_1 e^{w_-} - m_2 e^{- w_-} )
    \delta^{(p-1)}(\vec k_1 - \vec k_2)~,
\end{align}
where we have defined $w_{\pm} = \frac12 (w_1 \pm w_2)$. 
This expression reduces to
\begin{align}
   \langle \Psi^*_{m_2,l_2}  \Psi_{m_1,l_1} \rangle
    = \delta_{l_1 l_2} \delta(m_1 -m_2)
    \delta^{(p-1)}(\vec k_1 - \vec k_2) ~,
\end{align}
where the normalization has been set by the definition of the wave
function \eqref{wavefunction}. Note that there is no phase shift
in the two point function as in the case with a constant
$\Theta^{ij}$. The two point function in the closed string case
can be also computed using \eqref{NCcorrelator} with $\Theta^{ij}=0$,
thus giving the same result as in the open string one.

\subsection{Three point function}

It was shown in \cite{BCKR} that the three point function for
closed strings does not show any singular behavior since the main
contribution comes from the asymptotic regions. We anticipate that
the similar conclusion would be drawn also in open string case,
but the three point function may include non-trivial phase factor
due to the non-trivial flux. Using \eqref{NCcorrelator}, the three
point function can be computed as $(\tau_1 > \tau_2 > \tau_3)$
\begin{align}
 &\langle \Psi^*_{m_1,l_1} \Psi_{m_2,l_2} \Psi_{m_3,l_3} \rangle
  = \frac{\sqrt{m_1 m_2 m_3}}{\sqrt{2 \pi^3}}
  \int dw_1 dw_2 dw_3 e^{i(- l_1 w_1 + l_2 w_2 - l_3 w_3)}
    e^{\frac{i}{2}\theta m_2 m_3 \sinh (w_2 -w_3)}
   \nonumber \\ &\times
    \delta (-m_1 e^{w_1} + m_2 e^{w_2} + m_3 e^{w_3} )
    \delta (-m_1 e^{-w_1} + m_2 e^{-w_2} + m_3 e^{-w_3})
    \delta^{(p-1)}(-\vec k_1 + \vec k_2 + \vec k_3)~.
\end{align}
The phase shift has been simplified by using the delta functions as in
the two point function.
The integration w.r.t $w_1$ leads to
\begin{align}
 &\frac{\sqrt{m_1 m_2 m_3}}{\sqrt{2 \pi^3 }}
  \int dw_2 dw_3 e^{i(l_2 w_2 + l_3 w_3)}
    \frac{(m_2 e^{-w_2}+m_3 e^{-w_3})^{il_1-1}}
    {m_1^{il_1}}
    e^{\frac{i}{2}\theta m_2 m_3 \sinh (w_2 -w_3)}
   \nonumber \\ &\times
    \delta \left( - \frac{m_1^2}{m_2 e^{-w_2} + m_3 e^{- w_3}}
            + m_2 e^{w_2} + m_3 e^{w_3} \right)
    \delta^{(p-1)}(-\vec k_1 + \vec k_2 + \vec k_3)~.
\end{align}
Rewriting $w_{\pm}=\frac12(w_2 \pm w_3)$ and integrating
$w_+$ out, we find
\begin{align}
  &2 \sqrt{\frac{2 m_1 m_2 m_3}{\pi}} \delta_{l_1,l_2 + l_3}
  \int dw_- e^{i(l_2 - l_3) w_-}
    \frac{(m_2 e^{-w_-}+m_3 e^{w_-})^{il_1-1}}
    {m_1^{il_1}}
    e^{\frac{i}{2}\theta m_2 m_3 \sinh 2 w_-}
   \nonumber \\ &\times
    \delta \left(\frac{- m_1^2 + m_2^2 + m_3^3 + 2 m_2 m_3 \cosh 2 w_-}
                  {m_2 e^{-w_-}+m_3 e^{w_-}} \right)
    \delta^{(p-1)}(-\vec k_1 + \vec k_2 + \vec k_3)~.
\end{align}
Assuming $m_1 > m_2 + m_3$, we finally obtain by
integrating $w_-$ out as
\begin{align}
 \sqrt{\frac{2 m_1 m_2 m_3}{\pi}}
 \frac{\delta_{l_1,l_2 + l_3} (e^{i \xi_+} + e^{i \xi_-})}
   {M} ~,
\end{align}
where
\begin{align}
 &M =  \sqrt{m_1^2 - (m_2 + m_3)^2}\sqrt{m_1^2 - (m_2 - m_3)^2} ~,\nonumber \\
 &e^{i \xi_{\pm}} = e^{\pm \frac{i}{4}\theta M} e^{\pm i (l_2 - l_3) w_0}
   \left( \frac{m_2 e^{\mp w_0} + m_3 e^{\pm w_0}}{m_1} \right)^{il_1} ~,
  \\
 &\sinh  w_0 = \sqrt{\frac{m_1^2 - (m_2 + m_3)^2}{4m_2m_3}} ~. \nonumber
\end{align}
{}From the above result, we can see that the introduction of gauge
flux affects the amplitude in a non-trivial way, though the
amplitude is finite even without flux. Note that the shift due to
the non-commutativity is very small for the case when the mass
difference is very small $(m_1 \sim m_2 + m_2)$.

\subsection{Four point function}

We can compute four point function in the
same manner as before.
Both for closed and open strings, the $2 \to 2$ scattering
amplitudes are written in the form of\footnote{%
For closed strings we have to set $p=25$.}
\begin{align}
\begin{aligned}
  &\langle \Psi^*_{m_3,l_3} \Psi^*_{m_4,l_4}
           \Psi_{m_1,l_1} \Psi_{m_2,l_2} \rangle
  = \frac{\sqrt{m_1 m_2 m_3 m_4}}{\pi}
    \delta^{(p-1)}(\sum_{i=1}^4 \epsilon_i \vec k_i)
    \delta (\sum_{i=1}^4 \epsilon_i l_i) \\
  & \qquad \times \int dv_2 dv_3 dv_4 A(s,t,u)
    \delta (m_1 + \prod_{i=2}^4 \epsilon_i m_i v_i)
    \delta (m_1 + \prod_{i=2}^4 \epsilon_i m_i /v_i)
    \prod_{i=2}^4 v_i^{i \epsilon_i l_i -1}~,
    \label{4ptfunction}
\end{aligned}
\end{align}
where $\epsilon_1 = \epsilon_2 = - \epsilon_3 = - \epsilon_4 = 1$.
We have used $v_i=e^{w_i-w_1} \ (i=2,3,4)$ and the
Mandelstam variables
\begin{align}
s&=-(k_1+k_2)^2 = 2 m^2_{T}
  + m_1 m_2 (v_2 + \frac{1}{v_2}) - 2 \vec k_1 \cdot \vec k_2 ~,
  \nonumber\\
t&=-(k_1-k_3)^2 = 2 m^2_{T}
  - m_1 m_3 (v_3 + \frac{1}{v_3}) - 2 \vec k_1 \cdot \vec k_3 ~,
  \\
u&=-(k_1-k_4)^2 = 2 m^2_{T}
  - m_1 m_4 (v_4 + \frac{1}{v_4}) - 2 \vec k_1 \cdot \vec k_4 ~.
  \nonumber
\end{align}
The tachyon mass is $m^2_{T} = - 4/\alpha'$ for closed strings
and $m^2_{T}= - 1/\alpha ' $ for open strings.
Note that there is a non-trivial relation as $s+t+u = 4 m^2_T$.
For four point function, we cannot fix the position $z$ of one of the
four vertex operators, and $A(s,t,u)$ represents the contribution
coming from the integration of $z$. The amplitude is called as
Virasoro-Shapiro amplitude for closed strings and Veneziano
amplitude for open strings.

In the closed string case, Virasoro-Shapiro amplitude is defined as
\begin{align}
\begin{aligned}
 A_{VS} (s,t,u) &=
  \int d^2z |z|^{-\alpha ' u/2 -4} |1-z|^{-\alpha 't/2 - 4} \\
  &= 2 \pi \frac{\Gamma (-1- \frac{\alpha ' s}{4})
    \Gamma (-1 - \frac{\alpha ' t}{4})\Gamma (- 1 - \frac{\alpha ' u}{4}) }
      {\Gamma (2+ \frac{\alpha ' s}{4})
    \Gamma (2+ \frac{\alpha ' t}{4})\Gamma (2+ \frac{\alpha ' u}{4}) } ~.
\end{aligned}
\end{align}
Later we will use its high energy behavior. 
If we take $s,t,u$ very large,
then the amplitude is exponentially damping.
On the other hand, in the Regge limit, where $s \to \infty$ and $t$ is fixed,
the amplitude behaves as
\begin{align}
 A_{VS} \sim s^{2+\frac{\alpha' t}{2}} \frac{\Gamma (-1- \frac{\alpha ' t}{4})}
   { \Gamma (2+ \frac{\alpha ' t}{4})} ~.
\end{align}
The poles of the gamma function at $\alpha ' t = -4,0,4,\cdots$
correspond to the on-shell poles of exchanged states, and
the powers of $s$ are the spins of the states.
For example, the tachyon $m^2 = - 4/\alpha '$ has spin $0$,
the graviton $m^2 = 0$ has spin $2$, and so on.

Since we introduce non-trivial flux on the D-brane in open string case,
the Veneziano amplitude is modified due to the non-commutativity.
Taking care of the ordering of the four vertex operators and
making use of the momentum conservation, we can write the
amplitude as\footnote{Using the cyclic property of disk amplitude,
we set $\tau_1$ less than the other $\tau_i$. Then, the phase factor
involving $ m_1 $ and $w_1 $ becomes trivial due to the momentum
conservation. The other part of phase factor depends on the ordering of
$\tau_i$ because of the epsilon factor in \eqref{NCcorrelator}.}
\begin{align}
\begin{aligned}
 A_V (s,t,u) &= ( e^{ i (- \chi_{42} - \chi_{43} + \chi_{23})}
   + e^{i (- \chi_{42} - \chi_{43} + \chi_{23})} ) I (s,t) \\
    &+ ( e^{ i (- \chi_{42} + \chi_{43} - \chi_{23})}
   + e^{i (+ \chi_{42} - \chi_{43} + \chi_{23})} ) I (t,u) \\
    &+ ( e^{ i (- \chi_{24} + \chi_{34} + \chi_{23})}
   + e^{i (+ \chi_{42} - \chi_{43} - \chi_{23})} ) I (s,u) ~,
   \label{veneziano}
\end{aligned}
\end{align}
where
\begin{align}
 &I(x,y)=\int_0^1 dy y^{-\alpha ' x -2} (1-y)^{-\alpha ' t -2}
        = \frac{\Gamma(-\alpha ' x - 1)\Gamma(-\alpha ' y - 1)}
               {\Gamma(-\alpha ' x -\alpha ' y - 2)} ~,
\end{align}
and 
\begin{align}
\chi_{ij} = \frac{1}{4} m_i m_j 
 \left( \frac{v_i}{v_j} -  \frac{v_j}{v_i} \right) ~.
\label{ncphase}
\end{align}
The high energy behavior of Veneziano amplitude is very similar to
the Virasoro-Shapiro amplitude.
Again an interesting behavior is shown in the Regge limit as
\begin{align}
 I(s,t) \sim I(t,u) \sim s^{1+\alpha' t} \Gamma(-1-\alpha' t)
\end{align}
with exponentially damping $I(s,u)$.
The poles of exchanged states are at $\alpha ' t=
-1,0,1,\cdots$. The powers of $s$ are spins of the states, which
are $0$ for the tachyon $(m^2 = -1/\alpha')$, $1$ for gauge field
$(m^2 = 0)$, and so forth.

Now that we know the properties of the amplitude $A(s,t,u)$,
let us perform the integration of $v_i$ in \eqref{4ptfunction}.
Using the delta functions, the four point function can be
written as
\begin{align}
 &\frac{\sqrt{m_1 m_2 m_3 m_4}}{\pi }
    \delta^{(p-1)}(\sum_{i=1}^4 \epsilon_i \vec k_i)
    \delta (\sum_{i=1}^4 \epsilon_i l_i)
    \int dv_4 A(s,t,u)
    \frac{v_2^{il_2+1}v_3^{il_3+1}v_4^{il_4-1}}
    {m_2 m_3 |v_2^2 - v_3^2|} ~,
\end{align}
where $v_2$ and $v_3$ are replaced by
\begin{align}
 v_2 &= \frac{\alpha \beta + m_2^2 - m_3^2 \mp \sqrt{\Xi}}{2 m_2 \beta} ~,
 &v_3 &= - \frac{\alpha \beta + m_3^2 - m_2^2 \pm \sqrt{\Xi}}{2 m_3 \beta}
\end{align}
with
\begin{align}
 \alpha &= - m_1 + m_4 v_4 ~,
 &\beta &= - m_1 + \frac{m_4}{v_4} ~,
 &\Xi &= (m_2^2 - m_3^2)^2 - 2 \alpha \beta (m_2^2 + m_3^2) + \alpha ^2 \beta ^2 ~.
\end{align}
The integral w.r.t. $v_4$ may diverge at the high energy region
$v_4 \to \infty$. In $v_4 \to \infty$ limit, we find
\begin{align}
 v_2 &\sim \frac{m_4 v_4}{m_2} ~,
 &v_3 &\sim \frac{m_3}{m_1} ~,
 &t&\sim -(\vec k_1 - \vec k_3)^2 ~,
 &s&\sim m_1 m_4 v_4 ~,
\end{align}
therefore we can see that the limit corresponds to the Regge limit.

First, we examine the closed string case, which was already done
in \cite{BCKR}.\footnote{See \cite{LMS1,LMS2} for null brane
case.}
Using the Regge limit of Virasoro-Shapiro amplitude,
the four point function becomes
\begin{align}
 \frac{\Gamma(-1+\frac{\alpha ' (\vec k_1 - \vec k_3)^2}{4} )}
 {\Gamma(2-\frac{\alpha ' (\vec k_1 - \vec k_3)^2}{4} )}
  \int_{v_m}^{\infty} dv_4 v_4^{-\frac{\alpha '}{2}(\vec k_1 - \vec k_3)^2 + i(l_2 - l_4)} ~,
\end{align}
where we take a cut off $v_m \gg 1$. The integral diverges for
$(\vec k_1 - \vec k_3)^2 \leq \frac{2}{\alpha '}$ and $l_2 - l_4 =
0$, where graviton is exchanged. {}From this fact, it was claimed
in \cite{BCKR} that there is large back reaction due to the
graviton exchange near the big crunch/big bang singularity. Note
that there are other types of sources of divergences coming from
the poles of the gamma function. Since they are on-shell poles
of exchanged states, the divergence originates from the IR
effects, where this kind of divergence is usually regularized by the
$i\epsilon$-prescription. The UV divergence at $v_4 \to
\infty$ also comes from IR region $(\vec k_1 - \vec k_3)^2 \leq
\frac{2}{\alpha '}$, and hence this phenomenon may be regarded as
an example of IR/UV mixing \cite{LMS1,LMS2,BCKR}.

For our open string case, we can expect that the divergence
becomes milder since the spin in the contribution
$s^{\rm spin} \sim v_4^{\rm spin}$ is changed from spin$=2$
into spin$=1$, roughly speaking.
In fact, for open strings on D-brane without flux,
the four point function behaves as
\begin{align}
 \Gamma(-1 + \alpha ' (\vec k_1 - \vec k_3)^2 )
  \int_{v_m}^{\infty} dv_4 v_4^{-1 -\alpha ' (\vec k_1 - \vec k_3)^2
   + i(l_2 - l_4)} ~,
   \label{open4pt}
\end{align}
therefore the integral diverges only when
$(\vec k_1 - \vec k_3)^2 = 0$ and $l_2 - l_4 = 0$.
This is actually much better than in the closed string case,
because the singularity may be removed by
$i \epsilon$-prescription just like for the on-shell divergences,
which cannot be used to regularize the divergence in the closed string
amplitude.\footnote{Since the divergence arises at 
$t = 0$, the divergence may be regarded as IR/UV mixing as
in the closed string case (see also \cite{CSV}).
Moreover, we can show that the contribution of
the divergence comes from the region near the big crunch/big bang
singularity following \cite{BCKR}. 
However, we cannot identify the divergence as the
signal of the large back reaction since the open string tachyon does not
couple to dilaton or graviton in the tree level.
It would be interesting to examine the open string one loop effects, 
which may be interpreted as closed string exchanges in
string theory context.}

Finally, let us see whether the non-commutativity resolves the
singularity. Because the phase factor in \eqref{veneziano} behaves as
\begin{align}
 \chi_{42} &\sim {\cal O} (1) ~,
&\chi_{43} &\sim \chi_{23} \sim \frac{1}{4} m_1 m_2 v_4 ~,
\end{align}
only a constant factor is added to \eqref{open4pt}.
Therefore, we conclude that the non-commutativity does not
resolve the singularity in this case. 
This might be consistent with the
fact that only non-planer loop amplitudes would be regularized
by the introduction of non-commutativity (see, e.g., \cite{MRS}).

\section{Strings and D-branes in Grant space}

As discussed in \cite{LMS2,FM,HP},
the singular behavior of closed strings associated to the 
big crunch/big bang is largely improved by adding
non-trivial directions to the parabolic orbifold.
In our case, we introduce one more dimension $y$
and require the identification
\begin{align}
 x^{\pm} &\sim e^{\pm 2 \pi \gamma } x^{\pm} ~,
  &y &\sim y + 2 \pi R ~.
\end{align}
This space is called as Grant space \cite{Grant},
where the singular point at $x^+ = x^- =0$ disappears 
due to the translation along $y$ direction.
We should remark that the space still includes closed
time-like curves. In this section,
we examine how the resolution of the singularity affects
the (imaginary part of) annulus amplitudes for open strings
and open string scattering amplitudes.

\subsection{Torus amplitude}

The spectrum of closed strings in Grant space is almost
the same as the product of those for Misner space and
for a free boson $Y$. Only the difference is that the
conjugate momentum for $y$ direction is restricted as
$P_y = \frac{1}{R} (n - \gamma \hat J) $, where
$p_y = \frac12 (P_y + \frac{w R}{\alpha '})$ and
$\tilde p_y = \frac12 (P_y - \frac{w R}{\alpha '})$ for
left and right-movers, respectively.
Therefore, we can compute torus amplitude in the way similar
to the calculation in Misner space.
Since the amplitude was already obtained in \cite{CC}
in the oscillator formalism, we re-derive it in the 
path integral formalism.

As before, we concentrate on the twisted sector.
In the $(w,k)$-twisted sector, the boundary
condition for $Y$ is assigned as
\begin{align}
 Y ( \sigma_1 + 2 \pi , \sigma_2 ) &= 2 \pi R w + Y(\sigma_1 ,\sigma_2) ~,
&Y ( \sigma_1 , \sigma_2 + 2 \pi ) &= 2 \pi R k + Y(\sigma_1 ,\sigma_2)
\end{align}
along with \eqref{torusbc}, and the mode expansion is given by
\begin{align}
  Y (\sigma_1 , \sigma_2) = y_0 + R(w \sigma_1 + k \sigma_2)
   + \sum_{\{ m,n\} \neq \{ 0,0\} } a_{m,n} e^{i(m\sigma_1 + n\sigma_2)} ~.
\end{align}
While evaluating the path integral, a subtlety arises from
the overall normalization, which is fixed by following \cite{oneloop,Ginsparg}. 
Namely, we separate constant and
orthogonal parts as $Y(\sigma_i) = \tilde Y + Y'(\sigma_i)$ and
normalize as
\begin{align}
 \int {\cal D} \delta Y
e^{-\frac{1}{4 \pi \alpha} \int d^2\sigma \sqrt{g} (\delta Y)^2 } = 1 ~.
\end{align}
Therefore, we have
\begin{align}
 \int {\cal D} \delta Y'
  e^{- \frac{1}{4 \pi \alpha} \int d^2\sigma \sqrt{g} (\delta Y')^2 }
 = \left( \int d y e^{- \frac{1}{4 \pi \alpha}
    \int d^2\sigma \sqrt{g} y^2} \right)^{-1}
 = \left(\frac{\pi}{\frac{1}{4 \pi \alpha '} \int d ^2 \sigma \sqrt{g}}
   \right)^{-1/2}
 = \sqrt{\frac{\tau_2}{\alpha '}} ~.
\end{align}
Using this normalization, the partition function for the $Y$ part
is written as
\begin{align}
 {\cal T}^Y_{w,k} (\tau)
  = \sqrt{\frac{\tau_2}{\alpha '}}
   {{\rm Det} '}^{-1/2} (- \Delta)
 e^{- \frac{\pi R^2}{\alpha ' \tau_2} | k - \tau w |^2} ~,
\end{align}
where the contribution from non-zero modes is computed as
\begin{align}
 {\rm Det}' (- \Delta)
  = \prod_{\{m,n\}\neq\{0,0\}}
     \frac{1}{\tau^2_2} |n - \tau m|^2
  = 4 \pi^2 \tau^2_2 |\eta (\tau )|^4 ~.
\end{align}
Summing up all contributions, we obtain
\begin{align}
 {\cal T}_{w,k} (\tau)
  = \frac{1}{\sqrt{ 4 \pi^2 \alpha ' \tau_2}}
 \frac{e^{- \frac{\pi R^2}{\alpha ' \tau_2} | k - \tau w |^2
 - 2 \pi \tau_2 \gamma^2 w^2}}{|\vartheta_1 (i \gamma (k - \tau w) | \tau)|^2} ~,
\end{align}
which reproduces the result in \cite{CC}.%
\footnote{Notice that we have neglected total volume factor, i.e.,
$2 \pi R$ in this case.}
Adding extra 23 free bosons and ghost parts, we have the total
amplitude
\begin{align}
{\cal T}
 = \sum_{k,w} \int_{\cal F} \frac{d^2 \tau}{4 \pi^2 \alpha ' \tau_2^2}
  \frac{ e^{- \frac{\pi R^2}{\alpha ' \tau_2} | w - \tau k |^2
 - 2 \pi \tau_2 \gamma^2 w^2} }
   {(4\pi^2 \alpha ' \tau_2)^{11}|\vartheta_1 (i\gamma (w \tau +k)|\tau) \eta(\tau )^{21}|^{2}}  ~.
\end{align}
Notice that the amplitude receives suppression for
 large $|w|$ and/or $|k|$ from the new factor.

\subsection{Annulus amplitudes}

We have investigated D-branes in Misner space,
and found the divergent rates of open string pair creation and emission
of winding strings. Since singular behaviors of closed strings
are improved by introducing extra non-trivial direction in null brane 
case \cite{LMS2,FM,HP}, we expect that the string emission rates are
suppressed also in our case.
{}From this motivation, we will compute the string emission rates from
various D-branes in Grant space.
{}For simplicity, we assign Neumann or Dirichlet boundary condition
to $Y$, thus we have D0-brane, D1-brane and D2-brane.
Notice that there are two types of D1-branes which are
localized in the spatial direction of Misner space and in the
extra direction $y$.

We first examine the D$p$-brane localized in the spatial directions
of Grant space (D0-brane in Grant space part).
For the Misner space case, localized D$p$-brane has been 
constructed by summing over all image branes in the covering space,
and the same approach should be taken also for the Grant space case.
Let us assume that the original brane is located at $y=y_0$
in the covering space, then the $k$-th image brane is
at $y=y_0 + 2 \pi R k$. The open string stretched
between $k$-th and $k'$-th image branes has the boundary
conditions 
\begin{align}
\partial_{\tau} Y (\tau, 0) &= \partial_{\tau} Y (\tau, \pi) = 0 ~,
&Y(\tau, 0) &= y_0 + 2 \pi R k ~,
&Y(\tau, \pi) &= y_0 + 2 \pi R k' ~.
\end{align}
The mode expansion is
\begin{align}
 Y = y_0 + 2 \pi R k + 2R(k' - k)\sigma + i \sqrt{2 \alpha '}
 \sum_{n \in {\mathbb Z}, n \neq 0} \frac{\alpha^y_n}{n} e^{i n \tau} \sin n \sigma ~,
\end{align}
and the Virasoro generator is
\begin{align}
 L_0 = \frac{|k-k'|^2 R^2}{\alpha '}
 + \sum_{n > 0} \alpha ^y_{-n} \alpha^y_n ~.
\end{align}
Since the contribution to the annulus amplitude is computed as
\begin{align}
 {\rm Tr} e^{- 2 \pi t(L_0 - \frac{1}{24})}
  = \frac{e^{- \frac{2 \pi t |k-k'|^2 R^2}{\alpha '} }}{\eta (it)} ~,
\label{Dirichlet}
\end{align}
we obtain $(\nu = 2 \gamma (k - k'))$
\begin{align}
 {\cal A} = \sum_{k,k'} \int_0^{\infty} \frac{dt}{t}
  \frac{e^{-\pi t \nu^2 - \frac{2 \pi t |k-k'|^2 R^2}{\alpha '}}}
  {(8\pi^2 \alpha ' t)^{\frac{p}{2}}\vartheta_1 (t | \nu || i t) \eta (it)^{21}} ~.
\end{align}
Here we have added 23 extra boson contribution.
The imaginary part arises from the poles of integration by $t$ as
\begin{align}
 - 2 {\rm Im}\, {\cal A}
  = \sum_{k,k'}
   \sum_{n=1}^{\infty} \frac{(-1)^{n+1}}{ n (8 \pi ^2 \alpha ')^{\frac{p}{2}}}
    \left( \frac{|\nu|}{n} \right)^{\frac{p}{2}}
     \sum_{states} e^{- \frac{2 \pi n}{|\nu | } 
    ( \frac{R^2|k-k'|^2}{\alpha '} + N- 1) - \pi n |\nu| } ~.
\end{align}
This value is suppressed for large $|k - k'|$ from the new factor,
even though the rate is still divergent due the the Hagedorn density
$\rho_N \sim e^{2 \pi N}$ for $n/|\nu| < 1$. 
This suppression factor can be understood as follows.
The pair created open strings are stretched between image branes
also in $y$ direction, thus the open strings acquire extra mass and
hence the pair creation rate is suppressed.

Next we move to D1-brane extended to $y$ direction in Grant space.
In this case, we assign Neumann boundary condition to $Y$ field,
and this gives no changes from computation on the D$p$-brane in Misner
space case with $p \geq 1$. Therefore, the pair creation rate of open strings
is not altered from \eqref{paircreation}, and hence there is no
effect of introducing extra non-trivial direction.

In Grant space, there is another type of D1-brane,
which wraps the whole $x^{\pm}$ part but is localized in the $y$ 
direction. Since the brane does not wrap the whole Grant space,
only untwisted closed strings couple to the D-brane.
Thus the $X^{\pm}$ part is the same as the flat space case,
and $Y$ part is the same as D0-brane in Grant space, namely,
the all image branes at $y=y_0 + 2 \pi k R$ should be taken into
account. The annulus amplitude for the corresponding D$p$-brane
is then given as
\begin{align}
 {\cal A} = \sum_{k,k'} \int_0^{\infty} \frac{dt}{t}
  \frac{e^{- \frac{2 \pi t |k-k'|^2 R^2}{\alpha '}}}
  {(8\pi^2 \alpha ' t)^{\frac{p}{2}} \eta (it)^{24}} ~,
\end{align}
and it does not include imaginary part.
Therefore, we can say that the divergence of winding string emission 
is completely removed. 
This is because only winding closed strings would be emitted from the 
D-brane wrapping $x^{\pm}$ part as seen in section \ref{Dbrane}
but the winding strings do not couple to this type of D-brane.

Finally, let us analyze D2-brane wrapping the whole Grant space part.
In the path integral formulation, we have
boundary conditions for $Y$
\begin{align}
 Y(\sigma_1 , \sigma_2 + 2 \pi) &= 2 \pi R k + Y(\sigma_1 , \sigma_2 ) ~,
 &\partial_{\sigma_1} Y(\sigma_1=0,\pi , \sigma_2) &= 0
\end{align}
along with \eqref{D1bcpi1} and \eqref{D1bcpi2}, 
whose solution is
\begin{align}
 Y(\sigma_1 , \sigma_2 ) = y_0 + R k \sigma_2
 + \sum_{\{m,n\}\neq\{0,0\},m\geq 0}
 a^y_{m,n} \cos m \sigma_1 e^{i n \sigma_2} ~.
\end{align}
Therefore we multiply the Misner part by
\begin{align}
 {\cal A}_k^Y  = \frac{e^{-\frac{\pi R^2 k^2}{2 \alpha ' t}}}
 {(8 \pi ^2 \alpha ' t)^{\frac12} \eta (it)} = 
 \frac{e^{-\frac{\pi s R^2 k^2 }{2 \alpha ' }}}
      {(8 \pi ^2 \alpha ')^{\frac12} \eta (is)} ~.
\end{align}
In the oscillator formalism, we should take the twist operator
into account. Since the $Y$ satisfies the Neumann boundary condition,
the contribution is computed as
\begin{align}
 {\cal A}_k^Y
  &= {\rm Tr} (e^{2 \pi i R p_y})^k e^{- 2 \pi t (L_0 - \frac{1}{24})} \nonumber \\
  &= \int \frac{d p_y}{2 \pi}
   \frac{e^{2 \pi i k R p_y -2 \pi t (\alpha ' {p_y}^2 -\frac{1}{24})}}
   {\prod_{n > 0} (1-e^{- 2 \pi t n})}
  = \int \frac{d p_y}{2 \pi}
   \frac{e^{- 2 \pi t \alpha ' 
  \left( p_y - \frac{i R k}{2 t \alpha '} \right)^2 
 -\frac{\pi R^2 k^2}{2 \alpha ' t}
   +  \frac{\pi t}{12}}}
   {\prod_{n > 0} (1-e^{- 2 \pi t n})} ~,
\end{align}
which reproduces the path integral result.
Incorporating this part, we obtain the total cylinder amplitude as
$(\nu = \gamma k)$
\begin{align}
 {\cal A}
  = \sum_{k \neq 0} \int_0^{\infty} ds
    \frac{ e^{- \pi s \nu ^2 -\frac{\pi s R^2 k^2 }{2 \alpha ' }}}
   {2 \sinh (  \pi  | \nu| )
  (8 \pi^2 \alpha ')^{\frac{p-1}{2}}
     s^{\frac{25-p}{2}} \vartheta_1 ( | \nu |  s| is)\eta(is)^{21}} ~.
\end{align}
This leads to imaginary part from the poles of integration by $s$ as
\begin{align}
 - 2  {\rm Im}\, {\cal A}
  = \sum_{w \neq 0}
   \sum_{n=1}^{\infty} \frac{(-1)^{n+1}}{2 | \nu | \sinh (\pi | \nu |) (8 \pi ^2 \alpha ')^{\frac{p-1}{2}}}
    \left( \frac{|\nu|}{n} \right)^{\frac{25-p}{2}}
     \sum_{states} e^{- \frac{2 \pi n}{|\nu|} (\frac{R^2 k^2}{4 \alpha '} + N - 1 ) - \pi n |\nu|} ~.
\end{align}
Therefore, as in the D0-brane case, there is suppression from the new factor
for large $|k|$, even though the emission rate of winding strings is still
divergent for $n/|\nu| < 1$ due to the Hagedorn behavior $\rho_N \sim e^{2 \pi N}$.
This suppression comes from the fact that the closed strings
wrap the $y$ direction as well in the twisted sectors,
which gives extra mass to the emitted winding strings.

\subsection{Four point function}

As seen in section \ref{4pt}, the $2 \to 2$ scattering amplitude
for open strings diverges due to the big crunch/big bang
singularity.
Therefore, it is important to examine if the introduction of
extra direction resolves this singular behavior.
We use the D$p$-brane without flux wrapping the whole Grant space.
The wave function is defined as
 \begin{align}
 \Psi_{m,l,n,\vec p} =\sqrt{ \frac{m}{2 \pi R}}
  \int d w e^{\frac{i m }{\sqrt2}
   (X^- e^{- w} + X^+ e^{w}) + i w l + i p_y y + i \vec p \cdot \vec X}
 ~,
\end{align}
where $p_y = ( n - \gamma l )/R$.
A crucial difference from the wave function for the Misner space case
\eqref{wavefunction} is that now $l$
takes a real value, and hence the wave function behaves 
badly around $X^+ X^- \sim 0$ \cite{BCKR,Pioline3}. 
Therefore, it is more plausible to use a wave packet
\cite{Pioline3}\footnote{See \cite{LMS2} for null brane case.}
\begin{align}
  \Psi_{m,f(l),n,\vec p} = \int d l f(l)  \Psi_{m,l,n,\vec p}
\end{align}
with a smooth function of rapid decrease $f(l)$.
Then, the four point function of the wave packets is computed by%
\footnote{We assume $f^*(l) = f(-l)$ such that the formula would be simpler.}
\begin{align}
 \int d l_3 d l_4 dl_1 dl_2
  f^{(3)*}(l_3) f^{(4)*}(l_4) f^{(1)}(l_1) f^{(2)}(l_2)
  \langle \Psi^*_{m_3,l_3} \Psi^*_{m_4,l_4}
           \Psi_{m_1,l_1} \Psi_{m_2,l_2} \rangle ~.
\end{align}
Following the appendix of \cite{Pioline3}, the four point function
can be evaluated as $(v_4 = e^{\sigma})$
\begin{align}
 \int d \sigma e^{- \sigma \alpha ' |\vec p_1 - \vec p_3|^2}
  \left[ (\tilde f^{(1)} \tilde f^{(3)} (x)) *
        (\tilde f^{(2)*} \tilde f^{(4)*} (x) ) *
         \left({\textstyle \frac{1}{\sqrt\sigma}}
       e^{- \frac{x^2 R^2}{4 \sigma \alpha ' \gamma ^2}} \right) *
        \left(e^{- \frac{ |x| | \vec p_1 - \vec p_3 | R}{\gamma} } \right)
        \right] |_{x = \sigma}
\end{align}
in the large $v_4$ region.
We denote the Fourier transform of $f(l)$ as  $\tilde f(x)$
and the convolution product as $*$.
We have also set $n_1 = n_3$ in order to see the singular behavior.
Assuming that $f^{(i)} (l)$ decay faster enough, the four point function
is approximated as
\begin{align}
 \int d \sigma e^{- \sigma \left( \alpha ' |\vec p_1 - \vec p_3|^2
  + \frac{R^2}{4 \alpha ' \gamma ^2}
  + \frac{| \vec p_1 - \vec p_3 | R}{\gamma} \right)} ~.
\end{align}
We can see that the integral converges even around
$|\vec p_1 - \vec p_3 | = 0$.
Therefore, we conclude that the divergence
is removed by introduction of extra non-trivial direction
just like the null brane case \cite{LMS2,FM,HP}.
In fact, this is contrary to the closed string case where
the divergence on the four point function cannot be completely 
removed in the Grant space \cite{Pioline3}.
We should remark that we can reproduce the previous result
by taking the limit $R \to 0$.

\section{Conclusion} 
\label{conclusion}

In this paper we have studied various aspects of D-branes and
open strings on them in a space-time that contains big crunch/big
bang singularity. 
As a simple setup, we have chosen a Lorentzian orbifold 
${\mathbb R}^{1,1}/\Gamma$ with a discrete boost $\Gamma$,
which is known as Misner (Milne) space \cite{Misner}.
The construction is very simple thanks to the orbifold technique, 
but the structure is complex enough; 
The universe includes cosmological regions with big crunch and
big bang singularities and whisker regions with closed 
time-like curves (CTCs).
First, we have examined the spectrum of closed strings,
mainly, in the twisted sectors of the orbifold model,
where the winding strings can be of macroscopic size 
\cite{Pioline1,Pioline2,Pioline3,Pioline4}.
We have also computed the one-loop amplitude \cite{Nekrasov},
and shown the no-ghost theorem.
In particular, we have noticed that only transverse 
excitations contribute to the one-loop amplitude.

In the Misner space, there are D0-brane and D1-brane as in 
fig.~\ref{D0D1misner}. The D0-brane starts from the past infinity,
meets the big crunch singularity, passes through the regions with
CTC, and goes to the future infinity. The D1-brane wraps the
whole universe, and can include gauge flux, which induces
time-dependent non-commutativity in the effective theory.
We have computed the partition functions in terms of
both one-loop amplitudes for open strings and 
overlaps between boundary states.
Interestingly, we have observed the imaginary parts of the
amplitudes for both the D0-brane and the D1-brane,
which have been interpreted as follows.
In order to construct the D0-brane in the orbifold model,
we have to sum up the image branes generated by the discrete boost
in the covering space. The imaginary part of the annulus amplitude
corresponds to the absorption rate between image branes, and
this can be also interpreted as the pair creation rate
of open strings \cite{Bachas}.
For the D1-brane, the imaginary part is understood as the emission
of twisted closed strings. 
Since the emitted winding strings are localized in whisker regions,
the emission may be related to the chronology protection conjecture
\cite{Hagedorn}. We have extended this analysis into superstring case
to show that the cause of the D-brane instabilities is not related to 
the tachyonic state, which would be projected out in the superstring settings.

It was already known \cite{BCKR} that the $2 \to 2$ scattering 
amplitude for closed strings in Misner space is divergent
due to the graviton exchanges near the big crunch/big bang 
singularity. We have examined $2 \to 2$ scattering
for open strings on D-branes wrapping the Misner space part, and
found that the divergent structure is largely improved 
since gauge field exchanges instead of graviton.
On the D-brane wrapping the Misner space, we can include 
non-trivial gauge flux, and the flux
induces time-dependent non-commutativity. 
However, we have shown that the non-commutativity does not
resolve the divergence associated big crunch/big bang
singularity in the 4 point function.

Including an extra direction and taking orbifold by combining boost
and translation, we obtain a smooth manifold without 
big crunch/big bang singularity, which is  called as Grant space 
\cite{Grant}.
We have examined D-branes and open strings on them in Grant
space. The rates of open string pair creation and emission
of winding closed strings are mostly suppressed by introduction
of extra non-trivial direction. We have shown that
$2 \to 2$ open string scattering amplitude is also
regularized by the effect.

To conclude, we have introduced D-branes as probes to investigate
a big bang/big crunch universe, namely, the Misner space.
Undoubtfully, a most important problem is to study the back reaction
to this background. For example, we could not answer the question
what will happen after the end of open string pair creation
and emission of winding strings. A related question may be on the
closed string tachyon condensation in the twisted sectors 
\cite{Hagedorn,MS}.
The twisted tachyon arises for bosonic string or superstrings
with anti-periodic fermions. The tachyonic states
are naively obtained by setting 
$\omega^2 = \tilde \omega ^2 = - {\cal M}^2$ in \eqref{tomega},
which may lead to the classical trajectories
\begin{align}
X^{\pm} (\tau,\sigma) &= \pm
 \frac{\sqrt{2\alpha'} {\cal M}}{\nu} e^{\pm \nu \sigma} \sinh \nu \tau ~,
&X^{\pm} (\tau,\sigma ) &=
\frac{\sqrt{2\alpha'} {\cal M}}{\nu} e^{\pm \nu \sigma} \cosh \nu \tau ~,
\end{align}
for $\epsilon = \tilde \epsilon = 1$ and 
$\epsilon = - \tilde \epsilon = 1$, respectively.
These states are not localized around the big crunch/big
bang singularity, and are localized in the regions opposite to
the massive case. 
For Euclidean orbifold case \cite{APS},
the analysis of closed string tachyon becomes simpler, 
because the tachyon is localized at the fixed point. 
The analysis for Lorentzian orbifold case
is rather difficult due to the non-locality of
the twisted tachyon.

It is also interesting to apply this analysis
into time-dependent non-commutative field theories and AdS/CFT 
correspondence, for example, along the line of \cite{HS,DN,CLO}.
Since the effective theory on D-brane can be obtained by 
utilizing the orbifold methods, it might be possible to analyze
the theory even with time-dependent non-commutativity.
It is also worth to study open strings stretched between
different D-branes. In this case, the structure of poles in
moduli integral is changed, and hence the behavior of
emissions of open and/or closed strings would be different.
It might be important to study the scattering amplitude for
open strings on D0-branes, which should be similar to 
scattering amplitudes for twisted closed strings \cite{Pioline3}.

\subsection*{Acknowledgement}

We would like to thank Yu Nakayama, Makoto Natsuume, Soo-Jong Rey,
Yuji Sugawara and Atsushi Yamaguchi for useful discussions. 
RRN and KLP wish to thank high energy theory group 
Institute of Physics, India, especially Alok Kumar, 
for the hospitality where a part of the work was done. The
work of RRN and KLP was supported in part by INFN, by the
MIUR-COFIN contract 2003-023852, by the EU contracts
MRTN-CT-2004-503369 and MRTN-CT-2004-512194, by the INTAS contract
03-51-6346, and by the NATO grant PST.CLG.978785.


\appendix

\section{Theta and eta functions}
\label{formulae}

We have used the definition of theta functions as
\begin{align}
\begin{aligned}
 \vartheta_1 (\nu|\tau) &= 2 q^{\frac18} \sin \pi \nu
  \prod_{m=1}^{\infty} (1-q^m)(1-zq^m)(1-z^{-1}q^m) ~,
 \\
 \vartheta_2  (\nu|\tau) &= 2 q^{\frac18} \cos \pi \nu
  \prod_{m=1}^{\infty} (1-q^m)(1+zq^m)(1+z^{-1}q^m) ~,
 \\
 \vartheta_3  (\nu|\tau) &=
  \prod_{m=1}^{\infty} (1-q^m)(1+zq^{m-\frac12})(1+z^{-1}q^{m-\frac12}) ~,
 \\
 \vartheta_4  (\nu|\tau) &=
  \prod_{m=1}^{\infty} (1-q^m)(1-zq^{m-\frac12})(1-z^{-1}q^{m-\frac12}) ~,
\end{aligned}
\end{align}
where we have set $q=\exp ( 2 \pi i \tau )$ and $z = \exp (2 \pi i \nu )$.
Their modular transformations are
\begin{align}
\begin{aligned}
 \vartheta_1 \left(\frac{\nu}{\tau} \left| - \frac{1}{\tau} \right.\right)
  &= - i ( - i \tau )^{\frac{1}{2}}
    e^{ \frac{\pi i \nu ^2}{\tau} }
    \vartheta_1 ( \nu | \tau ) ~,
 &\vartheta_2 \left(\frac{\nu}{\tau} \left| - \frac{1}{\tau} \right.\right)
  &= ( - i \tau )^{\frac{1}{2}}
    e^ { \frac{\pi i \nu ^2}{\tau} }
    \vartheta_4 ( \nu | \tau ) ~, \\
 \vartheta_3 \left(\frac{\nu}{\tau} \left| - \frac{1}{\tau} \right.\right)
  &= ( - i \tau )^{\frac{1}{2}}
    e^{ \frac{\pi i \nu ^2}{\tau} }
    \vartheta_3 ( \nu | \tau ) ~,
 &\vartheta_4 \left(\frac{\nu}{\tau} \left| - \frac{1}{\tau} \right.\right)
  &= ( - i \tau )^{\frac{1}{2}}
    e^{ \frac{\pi i \nu ^2}{\tau} }
    \vartheta_2 ( \nu | \tau ) ~.
\end{aligned}
\end{align}
We have also used Dedekind eta function defined as
\begin{align}
 \eta (\tau) &= q^{\frac{1}{12}} \prod_{m=1}^{\infty} (1 - q^m) ~,
 &\eta \left( - \frac{1}{\tau} \right)
 &= ( - i \tau )^{\frac12 } \eta ( \tau ) ~.
\end{align}


\baselineskip=14pt

\end{document}